\documentclass[a4paper,11pt]{article}
\pdfoutput=1 

\usepackage{jheppub} 
\dedicated{This paper was published in a special issue of Geometric Mechanics dedicated to the memory of\\ Miguel Carlos {Mu\~noz}-Lecanda who, besides being a good friend, had a steady influence in the professional development of one os us (CB).}
\usepackage[T1]{fontenc} 
\subheader{Originally published in \textit{Geometric Mechanics}\\ https://doi.org/10.1142/S2972458925400015}

\newcommand{\dif}{\mbox{d}}

\usepackage{xcolor}
\usepackage{url}

\def\adst{AdS$_3$}
\def\adsd{AdS$_2$}
\def\adstsod{AdS$_3$/SO(2)}

\DeclareMathOperator{\arccosh}{arccosh}
\DeclareMathOperator{\arcsinh}{arcsinh}

\def\TCL{\tilde{\cal L}}
\def\TL{\tilde{L}}

\newcommand{\be}{\begin{equation}}
	\newcommand{\ee}{\end{equation}}
\newcommand{\bea}{\begin{eqnarray}}
	\newcommand{\eea}{\end{eqnarray}}
\newcommand{\bref}[1]{(\ref{#1})}
\newcommand{\nn}{\nonumber}

          \newcommand{\w}{\omega}
          
\newcommand{\h}{\eta}

\def\6{\partial} \def\7{\tilde} \def\8{\hat}

\def\CL{{\cal L}}

\def\={{\;=\;}}\def\+{{\;+\;}}

\def\ebox#1#2{\vskip 2mm{\vbox{\hrule\hbox{\vrule\kern3pt\vbox{\kern3pt
					{\begin{eqnarray}#1\label{#2}\end{eqnarray}}
					\kern3pt}\kern3pt\vrule}\hrule}}\vskip 2mm}
\def\tbox{\vskip 2mm{\vbox{\hrule\hbox{\vrule\kern3pt\vbox{\kern3pt
					{{\hfill {\small ${}^{notebook\; kiyoshi
								}$} \\
							\large \bf ~~\reptitle}\\ } 
				\kern3pt}\kern3pt\vrule}\hrule}}\vskip 2mm}

				\def\dif{{\rm d}}
				\def\deriv{\@ifnextchar[{\@deriv}{\@deriv[]}}
				\def\@deriv[#1]#2#3{\mathchoice%
{{\dif^{#1}#2\over\dif{#3}^{#1}}}{{\dif^{#1}#2/\dif{#3}^{#1}}}%
{{\dif^{#1}#2\over\dif{#3}^{#1}}}{{\dif^{#1}#2/\dif{#3}^{#1}}}}

\title{\boldmath  A general action for a particle in 2+1 anti-de Sitter space}

\author[a]{Carles Batlle,}
\author[b]{Joaquim Gomis}

\affiliation[a]{Departament de Matem\`atiques and IOC, 
	Universitat Polit\`ecnica de Catalunya\\
	EPSEVG, Av. V. Balaguer 1, E-08800 Vilanova i la Geltr\'u, Spain}
\affiliation[b]{Emeritus Professor of
	Universitat de Barcelona,
	Gran Via de les Corts Catalanes 585,
	08007 Barcelona,
	Spain}

\emailAdd{carles.batlle@upc.edu}
\emailAdd{joaquim.gomis@ub.edu}

\abstract{We discuss a general action for a particle in \adst\ using the non-linear realization framework. Critical sectors are found and characterized in terms of the parameters appearing in the Lagrangian, generalizing the known results for \adstsod.  A study of the dynamics of this system and of the geometrical meaning of the new terms in the action is also undertaken.

	Using two chiral copies of the general \adsd\ Lagrangian, we obtain an \adst\ Lagrangian which is the same, although expressed in different variables, that the general  \adst\ one, and from this one can identify the extra gauge transformations appearing in the critical sectors of the \adst\ theory. A differential equation connecting the \adsd$\times$\adsd\ chiral variables to the \adst\ covariant ones is obtained, and it is explicitly solved  at first order in the Goldstone bosons associated to boosts and rotations. 
	
}

\makeatletter
\gdef\@fpheader{}
\makeatother

\begin{document} 
	\maketitle
	\flushbottom

\section{Motivation and results}

Anti-de Sitter spaces have been the object of a renewed attention since the proposal of the AdS/CFT (anti-de Sitter/Conformal Field Theory) conjecture in 1997 \cite{Maldacena:1997re,Gubser:1998bc,Witten:1998qj}.
A review of the status of this remarkable conjecture can be found in \cite{Aharony:1999ti,Hubeny:2014bla}. Other applications of AdS spaces are presented in \cite{Gibbons:2011sg}, while the specific relevance of \adst\ in relation to black holes is discussed, for instance, in \cite{Kraus:2006wn} and
\cite{Martinec:2023zha}.

As explained in \cite{PhysRevD.48.1506,PhysRevD.79.105011}, properties of black holes in $2+1$ dimensions can be related to properties of point particles in an \adst\ background. 
In particular, understanding the dynamical sectors \cite{Alvarez:2007fw} of an \adst\ particle and their symmetries can be useful in order to study and classify  $2+1$ black holes and the Killing vectors that generate their isometries.

In a previous work \cite{BGKZ2014} we have studied the 
dynamics  of a spinning particle with arbitrary spin in \adst.
The action
was constructed from the coset
\adstsod, and it has two free parameters $M, J$, the mass and the spin of the particle \cite{Skagerstam:1989ti} \cite{deSousaGerbert:1990yp}.
The Lagrangian equations of motion revealed the presence of 
dynamical sectors depending on  the values of 
$M, J$ and $R$, the later being the \adst\ radius, which is considered fixed. It was found that there are three sectors depending on the values of 
$\kappa=J^2-M^2R^2$.
For the subcritical  sector ($\kappa <0$) and supercritical sector
($\kappa >0$) cases, it was seen that, in spite of the particle having arbitrary spin $J$, the equations of motion coincide with those of  the geodesics of a spinless particle in \adst. For the critical case  $(\kappa=0)$
there exists an extra gauge transformation which further reduces the physical degrees of freedom. This additional gauge transformation is the bosonic analog of the kappa
symmetry  for superparticles \cite{deAzcarraga:1982dw,Siegel:1983hh}. The orbits correspond in this case to the  geodesics in \adsd. 
The presence of sectors in the motion of the particle with arbitrary spin $J$ in \adst\ is in correspondence with the spectrum of a BTZ black hole \cite{MZ} 
with mass $M$ and angular momentum $J$. In particular, the extremal black hole corresponds to the critical 
sector $\kappa=0$.

In this paper we will extend the results in \cite{BGKZ2014} and write the most general particle action with the lowest order of derivatives in  terms of the space-time and Goldstone boson variables.
The Lagrangian will be constructed
by considering the  trivial coset
$\text{\adst}/1$ in the non-linear realization approach \cite{Gomis:2006xw}\cite{Gomis:2013NPB}\cite{Bergshoeff:2022eog}.
The action is written in terms of the world-line pull-backs of the six components of the  Maurer-Cartan 1-form. It contains six free parameters, four more than the Lagrangian constructed in \cite{BGKZ2014} from the coset \adstsod.
As we will see, there are sectors in the dynamics of this particle, and in particular 
critical sectors, for which there is an extra gauge transformation, besides the diffeomorphism of the worldline and the SO(2) local rotation, 
that further reduces the number of  physical degrees of freedom. We study the equations of motion in the non-critical sectors, and  write down the modifications to the geodesic equations due to the new terms.

Inspired by  the Lie algebra isomorphism $so(2,1)\times so(2,1)\sim so(2,2)$, we will study a general Lagrangian in \adsd\ and, after eliminating the non-dynamical variable
corresponding to the Goldstone boson associated to the \adsd\ boost, the resulting Lagrangian contains a redefined mass term together with an interaction with an electromagnetic background \cite{Anabalon:2006ii}.
Using two chiral copies of this \adsd\ Lagrangian, we obtain an \adst\ Lagrangian which is the same, although expressed in different variables, that
the general  \adst. This yields a further understanding of the critical sectors of the \adst\ theory, which turn out to correspond to the switching off of one of the chiral copies.

A differential equation connecting the \adsd$\times$\adsd\ chiral variables to the \adst\ covariant ones is obtained, and it is explicitly solved  at first order in the Goldstone bosons associated to boosts and rotations.

The paper is organized as follows. In Section \ref{sec_AdS3_lag} we construct
the most general \adst\ particle Lagrangian using the nonlinear realization method. The appearance of critical sectors and the form of the equations of motion are investigated. A  geometrical interpretation of this action is presented in Section \ref{sec_AdS3_geo}.  In Section \ref{sec_AdS2_lag} we obtain the most general \adsd\ Lagrangian and in Section \ref{sec_AdS2xAdS2_lag}, taking two chiral copies of it, we relate it to the  \adst\ Lagrangian constructed in Section \ref{sec_AdS3_lag}. The map between the chiral and covariant coordinates is studied in Section \ref{sec_map}, and the resulting differential equation is solved to first order in the boost and rotation coordinates.  Section \ref{sec_diss} discusses the results and some possible improvements. \ref{apMC} contains explicit expressions for the Maurer-Cartan forms for AdS$_3$/SO(2) and AdS$_3$. \ref{apAdS2} and \ref{EOM_L0} present a study of the symmetries of the general AdS$_2$ Lagrangian and of the solution of the equations of motion obtained from its massive part, respectively. Finally, a closed form for the embedding  of \adstsod\ into  AdS$_2\times$AdS$_2$ is obtained in \ref{embedding}.

In this paper, the tangent space metric is $\h_{ab}=\text{diag} (-++)$ and we use the following conventions: $\epsilon^{012}=+1$, indices $m, n, \ldots$ refer to  the spacetime manifold where the particle moves, and $a,b,\ldots=0,1,2$ are tangent spacetime indices, with primed indices restricted to spatial directions, $a',b',\ldots=1,2$.


\section{ A general \adst\ Lagrangian} 
\label{sec_AdS3_lag}

The non-linear realization technique was originally proposed by Weinberg \cite{PhysRev.166.1568} and by Coleman, Wess and Zumino \cite{PhysRev.177.2239} in relation to the chiral symmetry of effective actions in quantum field theory. The central idea of the method was extended in \cite{BANDOS199277}\cite{Bandos:1995zw} and in \cite{West:2000hr}\cite{Gomis:2006xw} to construct actions for relativistic (super)branes having a given space-time symmetry group, and was later applied to the construction of the action of other point-like and extended systems with prescribed relativistic and non-relativistic symmetries. Section 5 of \cite{Bergshoeff:2022eog} contains a detailed exposition of the method. In this paper the non-linear realization approach will be applied to the construction of particle actions with anti-de Sitter symmetry in $1+1$ and $2+1$ space-times.

The Lie algebra associated to the group AdS$_3$ is isomorphic to $so(2,2)$, with generators $P_a$, $M_{ab}$,
\begin{eqnarray}\label{cba2}
	\left[P_a,P_b\right]&=&-i \frac{1}{R^2}M_{ab},\;\left[P_{a},M_{cd}\right]=-i\h_{a[c}P_{d]}, \nn\\ 
	\left[M_{ab},M_{cd}\right]&=&-i\h_{b[c}M_{ad]}+i\h_{a[c}M_{bd]}.
\end{eqnarray}

We locally parametrize $g\in \text{AdS}_3$ as
\be
g=g_{\frac{AdS_3}{SO(2)}} e^{i M_{12}\phi }
\ee
where $g_{\frac{AdS_3}{SO(2)}}$ is an element of AdS$_3$/SO(2),
\begin{equation}\label{cba_new_1}
	g_{\frac{AdS_3}{SO(2)}}=e^{iP_0 x^0}e^{i P_1 x^1}e^{i P_2 x^2} e^{i M_{02} v^1} e^{-i M_{01} v^2}.
\end{equation}

The Maurer-Cartan (MC) form $\Omega=-ig^{-1}\dif g$ verifies the equation $d\Omega+i \Omega\wedge\Omega=0$
and 
can be expanded in terms of its components as\footnote{We use $\TL$ in order to distinguish these components from those of \adstsod, which will also appear.}
\begin{equation}
	\Omega=  P_a \TL^a + \frac{1}{2} M_{ab}\TL^{ab}
	=\TL^0 P_0 + \TL^1 P_1 + \TL^2 P_2 + \TL^{01} M_{01}+ \TL^{02} M_{02}+ \TL^{12} M_{12}. \label{cba_N1}
\end{equation}
Explicit expressions of the $1-$forms\footnote{Components of the MC form associated to a group element are denoted generically by $L^A$, with $A$ in the set $\{0,1,2,01,02,12\}$, while the corresponding  world-line pull-back is written as $\CL^A$.}
$\TL^A$
corresponding to a given  parametrization of $g$ are given in \ref{apMC}.

The MC forms $\TL^A$, $A=0,1,2,01,02,12$ 
can be written as 
\begin{eqnarray}
	\TL^a &=& e^b \Phi_b^{\ a}{ (v^{a'},\phi)},\quad a=0,1,2, \label{MC1}\\
	\TL^{ab} &=& \Phi_c^{\ a}{ (v^{a'},\phi)} \left( \eta^{cd} \dif + \omega^{cd}\right)\Phi_d^{\ b}{ (v^{a'},\phi)},\quad a,b=0,1,2,\ b>a,\label{MC2}
\end{eqnarray}
where $\Phi_b^{\ a}{(v^{a'},\phi)}$  is the Lorentz transformation
\be\label{PhiRep}
\Phi_b^{\ a}{  (v^{a'},\phi)}=\!
\begin{pmatrix} \cosh v^1& 0&-\sinh v^1\cr 0 & 1& 0 \cr -\sinh v^1&0 &\cosh v^1 \end{pmatrix}\! \!\!
\begin{pmatrix} \cosh v^2&\sinh v^2& 0 \cr \sinh v^2 &\cosh v^2& 0 \cr 0 & 0 &1 \end{pmatrix} \! \!\!
\begin{pmatrix} 1 & 0 & 0 \cr0 &  \cos\phi & \sin\phi \cr 0 &  -\sin\phi & \cos\phi \end{pmatrix}
\ee
and $e^a, \omega^{ab}$ are the vielbein and the spin connection of AdS$_3$.
In the flat limit $R\to\infty$,  $\omega^{ab}=0$.

A general AdS$_3$ invariant Lagrangian with lowest order in the derivatives is given by
\begin{equation}\label{cba_new_26_b}
	\CL = - M \TCL^0 - J \TCL^{12}  + A_1 \TCL^1 + A_2 \TCL^2  + B_1 \TCL^{01} + B_2 \TCL^{02}, 
\end{equation}
where the parameters $M$, $A_1$ and $A_2$ have mass dimension $1$, while  $J$, $B_1$ and $B_2$ are dimensionless, and where $\TCL^A \dif\tau$ is the pull-back of the MC form $\TL^A$ on the world-line parametrized by $\tau$.

In order to write the equations of motion, it is useful to consider the variation of the MC forms under a general variation of the coordinates.
If we denote by $Z^A$ the coordinates of an arbitrary parametrization of a generic group with Lie algebra $[G_A,G_B]=i f_{AB}^{\ \ C}G_C$,
we have 
\cite{Anabalon:2006ii}
\begin{equation}\label{cba_N3}
	\delta L^A = \dif [\delta Z]^A + \frac{1}{2} f_{BC}^{\ \ A}(L^C [\delta Z]^B)- [\delta Z]^C L^B.
\end{equation} 
with $[\delta Z]^A = \delta Z^B L_B^{\ A}$, and $L_B^{\ A}$ is
defined by $L^A= \dif Z^B L_B^{\ A}$.

Using the structure constants from (\ref{cba2}) one gets, for the variation of (\ref{cba_new_26_b}),
\begin{eqnarray}
	\delta\CL &=& -M\frac{\dif}{\dif \tau} [\delta Z]^0 +
	\frac{1}{2}[\delta Z]^0 \left( A_1 \TCL^{01} + A_2 \TCL^{02} - B_1/R^2\ \TCL^1 - B_2/R^2\ \TCL^2            
	\right)\nn\\
	&+&  A_1\frac{\dif}{\dif \tau} [\delta Z]^1 +
	\frac{1}{2}[\delta Z]^1 \left( -M \TCL^{01} - A_2 \TCL^{12} + B_1/R^2\ \TCL^0 + J/R^2\ \TCL^2          
	\right)\nn\\
	&+& A_2 \frac{\dif}{\dif \tau} [\delta Z]^2 +
	\frac{1}{2}[\delta Z]^2 \left( -M \TCL^{02} + A_1 \TCL^{12} + B_2/R^2\ \TCL^0 - J/R^2\ \TCL^1            
	\right)\nn\\
	&+& B_1\frac{\dif}{\dif \tau} [\delta Z]^{01} +
	\frac{1}{2}[\delta Z]^{01} \left( M \TCL^{1} - A_1 \TCL^{0} - B_2 \TCL^{12} + J \TCL^{02}            
	\right)\nn\\
	&+& B_2\frac{\dif}{\dif \tau} [\delta Z]^{02} +
	\frac{1}{2}[\delta Z]^{02} \left( M \TCL^{2} - A_2 \TCL^{0} + B_1 \TCL^{12} - J \TCL^{01}            
	\right)\nn\\
	&-& J\frac{\dif}{\dif \tau} [\delta Z]^{12} +
	\frac{1}{2}[\delta Z]^{12} \left( -A_1 \TCL^{2} + A_2 \TCL^{1} - B_1 \TCL^{02} + B_2 \TCL^{01}            
	\right).\label{cba_N5}
\end{eqnarray}

It can be seen that, for a given parametrization of the group,  the six $[\delta Z]^A$ are independent linear combinations of the six variations of $x^a$, $a=0,1,2$, $v^{a'}$, $a'=1,2$, and $\phi$, and hence they can be put independently to zero.

If we denote by $[\delta Z]$ the row vector with components $[\delta Z]^A$ and by $V_\CL$ the column vector with $\TCL^A$, the variation of $\delta\CL$ can be written, 
up to total derivative terms, as
\begin{equation}\label{cba_N7} 
	\delta\CL = \frac{1}{2}[\delta Z] S\  V_{\CL},
\end{equation} 
where S is given by
\begin{equation}\label{cba_N6}
	S=\begin{pmatrix}
		0 & -B_1/R^2 & -B_2/R^2 & A_1 & A_2 & 0 \\
		B_1/R^2 & 0 & J/R^2 & -M & 0 & -A_2  \\
		B_2/R^2 & -J/R^2 & 0 & 0 & -M & A_1 \\
		-A_1 & M &  0 & 0 & J & -B_2 \\
		-A_2 & 0 & M & -J & 0 & B_1 \\
		0 & A_2 & - A_1 & B_2 & -B_1 & 0
	\end{pmatrix}.
\end{equation}

A detailed study shows that the rank of $S$ is, at most, 4, which corresponds to non-critical sectors, and the rank drops to 2 for the two  cases given by plus and minus signs in the following relations:
\begin{eqnarray}
	J &=& \pm MR,\label{JMR}\\
	B_1 &=& \mp R A_2,\label{BRA2}\\
	B_2 &=& \pm R A_1,\label{BRA1}
\end{eqnarray}
which shows the existence of two critical sectors.
Notice that the rank cannot be less than two unless $J=M=A_1=A_2=B_1=B_2=0$, in which case there is no Lagrangian.

The fact that the rank is not maximal  reflects the existence of gauge transformations. In the non-critical sectors we have two transformations, the diffeomorphisms of the world line and local SO(2) rotations, while for  the critical sectors, due to the drop of the rank from $4$ to $2$, one has two more gauge transformations, yielding a total of four. 
These extra gauge transformations, together with the corresponding equations of motion, will be discussed in Section \ref{sec_AdS2xAdS2_lag} in terms of the \adsd\ Lagrangians.

In order to find  the gauge  transformations we impose
$  [\delta Z] S=0$ or, using that $S^T=-S$, which is a consequence of the skew-symmetry of the structure constants of the Lie algebra, $S\ [\delta Z]^T=0$. Hence
\begin{equation}\label{cba_N8}
	[\delta Z] \ \text{yields a gauge transformation}\ \Leftrightarrow \ [\delta Z]^T\in \ker S.
\end{equation}
Since in the non-critical cases $\ker S$ is two-dimensional, a general gauge transformation can be expressed in terms of two arbitrary $[\delta Z]^A$, 
for instance $[\delta Z]^0$ and $[\delta Z]^{12}$.   An explicit basis for $\ker S$ is given by $\{S_1,S_2\}$, with
\begin{align}
	S_1&=(MR^2, A_1 R^2, A_2 R^2, B_1, B_2 , J)^T,\nonumber\\
	S_2&=(J,B_2, -B_1, -A_2, A_1, M)^T.
	\label{kerS}
\end{align}

If we denote by $L$ the matrix with elements $L_A^{\ B}$ then, using $[\delta Z]^A = \delta Z^B L_B^{\ A}$ one has that an explicit form for the gauge transformation of the state variables 
$Z^A$ is given by
\begin{equation}
	(\delta_{\text{gauge}} Z)^T = L^{-T} \left(\epsilon_1(\tau) S_1 + \epsilon_2(\tau) S_2 \right),  
	\label{explicitGauge}	
\end{equation} 
with $\epsilon_{1,2}$ arbitrary functions of $\tau$.

The four independent equations of motion in the non-critical sectors are given by four independent rows  of $S V_{\CL}=0$, which can be selected as 
\begin{eqnarray}
	\TCL^1 &=& \frac{1}{J^2-M^2R^2}\left((B_2J-A_1MR^2)\TCL^0 + R^2(A_1J-B_2 M)\TCL^{12}\right),\label{EOM1}\\
	\TCL^2 &=&  \frac{1}{J^2-M^2R^2}\left(-(B_1J+A_2MR^2)\TCL^0 + R^2(A_2J+B_1 M)\TCL^{12}\right),\label{EOM2}\\
	\TCL^{01} &=& \frac{1}{J^2-M^2R^2}\left(-(A_2J+B_1M)\TCL^0 + (B_1J+ A_2 MR^2)\TCL^{12}\right),\label{EOM01}\\
	\TCL^{02} &=& \frac{1}{J^2-M^2R^2}\left((A_1J-B_2M)\TCL^0 + (B_2J- A_1 MR^2)\TCL^{12}\right).\label{EOM02}
\end{eqnarray}

Using the world-line pull-backs of (\ref{MC1}) and (\ref{MC2}), we can write 
\begin{eqnarray}
	\TCL^a &=& \dot x^m e_m^{\ b} \Phi_{b}^{\ a}{ (v^{a'},\phi)},\label{EOM-a}\\
	\TCL^{ab} &=& \Phi_{c}^{\ a} \eta^{cd}{ (v^{a'},\phi)} \dot \Phi_{d}^{\ b}{ (v^{a'},\phi)} + \Phi_c^{\ a}{ (v^{a'},\phi)} \dot x^m  \omega_m^{\ cd}\Phi_d^{\ b}{ (v^{a'},\phi)}.\label{EOM-ab}
\end{eqnarray}

Equations (\ref{EOM1})---(\ref{EOM02}) can be written as
\begin{eqnarray}
	\TCL^{a'} &=& \alpha^{a'} \TCL^0 + \beta^{a'}\TCL^{12},\ a'=1,2,\label{EOMa}\\
	\TCL^{0a'} &=& \gamma^{a'} \TCL^0 + \delta^{a'}\TCL^{12}, \ a'=1,2,\label{EOM0a}
\end{eqnarray}
where the constant coefficients $\alpha^{a'}$, $\beta^{a'}$, $\gamma^{a'}$ and $\delta^{a'}$ can be read off from (\ref{EOM1})---(\ref{EOM02}). Using (\ref{EOM-a}) and (\ref{EOM-ab}), equation (\ref{EOMa}) becomes
\begin{eqnarray}
	\dot x^m e_m^{\ b}\Phi_b^{\ a'}{ (v^{a'},\phi)} &=& \alpha^{a'} \dot x^m e_m^{\ b} \Phi_b^{\ 0}(v^{a'},\phi)\nonumber\\
	&+& \beta^{a'} \left( \Phi_c^{\ 1}{ (v^{a'},\phi)}\eta^{cd} \dot{\Phi}_d^{\ 2}{(v^{a'},\phi)} + \Phi_c^{\ 1}{ (v^{a'},\phi)} \dot x^m \omega_m^{\ \ cd}\ \Phi_d^{  \ 2}  
	{ (v^{a'},\phi)}\right).\nonumber\\
	& &
	\label{report1}
\end{eqnarray}

One can try a perturbative solution of (\ref{report1}) in $\alpha^{a'}$, $\beta^{a'}$, starting with the zeroth order solution
\begin{equation}\label{old-a'}
	\Phi_a^{\ 0} {}^{(0)}{ (v^{a'},\phi)} = \frac{1}{\sqrt{-g}} \dot x^m e_m^{\ b} \eta_{ba},
\end{equation}
where 
\begin{equation}\label{cba-g}
	g = \dot x^n e_n^{\ c}\ \eta_{cb}\ \dot x^m e_m^{\ b} < 0
\end{equation}
is the induced world-line metric. At fist order we write
\begin{equation}\label{new-a'}
	\Phi_a^{\ 0}{}^{(1)} = \frac{1}{\sqrt{-g}} \dot x^m e_m^{\ b} \eta_{ba} + K_a,
\end{equation}
with $K_a$ a linear combination of  $\alpha^{a'}$, $\beta^{a'}$ to be determined. This can be manipulated to yield
\begin{equation}\label{2new-a'}
	\dot x^m = \sqrt{-g} \Phi_a^{\ 0}{}^{(1)} e_n^{\ a} g^{nm} - \sqrt{-g} K_a  e_n^{\ a} g^{nm},
\end{equation}
and (\ref{report1}) becomes then, to first order in $\alpha^{a'}$, $\beta^{a'}$,
\begin{eqnarray}
	\lefteqn{\sqrt{-g} \Phi_a^{\ 0}{}^{(0)}
		\eta^{ab} \Phi_b^{a'}{}^{(0)} - \sqrt{-g} K_a \eta^{ab} \Phi_b^{a'}{}^{(0)} =}\nonumber\\
	&-& \alpha^{a'} \sqrt{-g} + \beta^{a'} \left( \Phi_c^{\ 1}{}^{(0)}\eta^{cd} \dot{\Phi}_d^{\ 2}{}^{(0)} + \Phi_c^{\ 1}{}^{(0)} \dot x^m \omega_m^{\ \ cd}\ \Phi_d^{  \ 2}{}^{(0)}\right).
	\label{report2}
\end{eqnarray}
The first term in the left-hand side is zero due to $\eta^{0a'}=0$, and one gets then
\begin{equation}
	K_a \eta^{ab} \Phi_b^{a'}{}^{(0)} = \alpha^{a'} -\frac{1}{\sqrt{-g}}\beta^{a'} \left( \Phi_c^{\ 1}
	{}^{(0)}\eta^{cd} \dot{\Phi}_d^{\ 2}{}^{(0)} + \Phi_c^{\ 1}{}^{(0)} \dot x^m \omega_m^{\ \ cd}\ \Phi_d^{  \ 2}{}^{(0)}\right).
	\label{report3}
\end{equation} 
These are two equations ($a'=1,2$) for the three unknowns $K_a$, $a=0,1,2$. However, the condition $\Phi_c^{\ 0}{}^{(0)}\eta^{ca}\Phi_a^{\ 0}{}^{(0)}=-1$ applied to (\ref{new-a'}) imposes, to first order in $K_a$,
\begin{equation}
	\dot x^m e_m^{\ b} K_b =0,
	\label{report4}
\end{equation}
which, at first order in the perturbation parameters, is equivalent to
\begin{equation}
	\Phi_b^{\ 0}{}^{(0)} \eta^{ba} K_a =0.
	\label{report4b}
\end{equation}
Equations (\ref{report3}) and (\ref{report4b}) can be written together as
\begin{equation}
	\Phi_b^{\ c}{}^{(0)} \eta^{ba} K_a =T^c, 
	\label{report4c}
\end{equation}
with $T^0\equiv 0$ and $T^{a'}$ given by the right-hand side of (\ref{report3}).  From (\ref{report4c}) one has
\begin{equation}
	\label{report4c_sol}
	K_a = \Phi_a^{\ b}{}^{(0)}\eta_{bc}T^c=\Phi_a^{\ b'}\eta_{b'c'}T^{c'},
\end{equation}
and the three $K_a$ and $\Phi^1{}^{(0)}$, $\Phi^2{}^{(0)}$
are fully determined in terms of $\dot x$, $e$, $\omega$

Using $\TCL^{a'0}=-\TCL^{0a'}$, the remaining equations of motion (\ref{EOM0a}) are
\begin{eqnarray}\label{report5}
	\lefteqn{\Phi_c^{a'} \eta^{cd}\dot{\Phi}_d^{\ 0} + \Phi_c^{\ a'} \dot x^m \omega_m^{\ \ cd}\Phi_d^{\ 0}  =}\nonumber\\
	&-&\gamma^{a'} \dot x^m e_m^{\ b} \Phi_b^{\ 0} - \delta^{a'} \left( \Phi_c^{\ 1}\eta^{cd} \dot{\Phi}_d^{\ 2} + \Phi_c^{\ 1} \dot x^m \omega_m^{\ \ cd}\ \Phi_d^{  \ 2}\right).
\end{eqnarray}
Since $\TCL^{00}=0$, if we define $\gamma^0=\delta^0=0$, we can extend $a'$ to $a$ in the above equation, and acting then with $\Phi_f^{\ e}\eta_{ea}$ one gets
\begin{eqnarray}
	\label{report6}
	\lefteqn{\dot{\Phi}_f^{\ 0} + \eta_{fc}\dot x^m \omega_m^{\ \ cd} \Phi_d^{\ 0} =}\nonumber\\ &-&\gamma^{a} \Phi_f^e \eta_{ea}\dot x^m e_{m}^{\ b}\Phi_b^{\ 0}
	-\delta^a \Phi_f^{\ e} \eta_{ea} \left( \Phi_c^{\ 1}\eta^{cd} \dot{\Phi}_d^{\ 2} + \Phi_c^{\ 1} \dot x^m \omega_m^{\ \ cd}\ \Phi_d^{  \ 2}\right).
\end{eqnarray}
Using (\ref{new-a'}) to zeroth order, the first term on the right-hand side becomes
$$
\gamma^a \Phi_f^{\ e} \eta_{ea} \sqrt{-g},
$$ 
while the zeroth order contribution from the first term in the left-hand side is
$$
e_n^{\ c}\eta_{cf} G^n
$$
where
\begin{equation}\label{geo}
	G^n = \frac{\dif}{\dif \tau}\frac{\dot x^n}{\sqrt{-g}} + \Gamma_{ml}^n   \frac{\dot x^m \dot x^l}{\sqrt{-g}}
\end{equation}
is the standard geodesic term. Adding the remaining first order terms, one gets the equations of motion (EOM) 
\begin{eqnarray}
	\lefteqn{e_n^{\ b}\eta_{ba}\left(    
		\frac{\dif}{\dif \tau}\frac{\dot x^n}{\sqrt{-g}} + \Gamma_{ml}^n   \frac{\dot x^m \dot x^l}{\sqrt{-g}}
		\right) =}\nonumber\\ & & \eta_{ab} \dot x^m \omega_m^{\ \ bc} K_c -  \frac{\dif}{\dif \tau} K_a 
	+ \gamma^b \Phi_a^{\ c}\eta_{cb} \sqrt{-g} \nonumber\\
	& & - \delta^b \Phi_{a}^{\ c }\eta_{cb} \left( \Phi_d^{\ 1}\eta^{de} \dot{\Phi}_e^{\ 2} + \Phi_d^{\ 1} \dot x^m \omega_m^{\ \ de}\ \Phi_e^{  \ 2}\right),
	\label{report7}
\end{eqnarray}
which, taking into account $\gamma^0=\delta^0=0$, boils down to
\begin{eqnarray}
	\lefteqn{e_n^{\ b}\eta_{ba}\left(    
		\frac{\dif}{\dif \tau}\frac{\dot x^n}{\sqrt{-g}} + \Gamma_{ml}^n   \frac{\dot x^m \dot x^l}{\sqrt{-g}}
		\right)}\nonumber\\ &=& \eta_{ab} \dot x^m \omega_m^{\ \ bc} K_c -  \frac{\dif}{\dif \tau} K_a \nonumber\\
	& & + \gamma^{b'} \Phi_a^{\ c'}\eta_{c'b'} \sqrt{-g} \nonumber\\
	& & - \delta^{b'} \Phi_{a}^{\ c' }\eta_{c'b'} \left( \Phi_d^{\ 1}\eta^{de} \dot{\Phi}_e^{\ 2} + \Phi_d^{\ 1} \dot x^m \omega_m^{\ \ de}\ \Phi_e^{  \ 2}\right).
	\label{report8}
\end{eqnarray}
Once $\Phi^1$, and hence $\Phi^2$, is gauge fixed and (\ref{report4c}) is solved for $K_a$, the right-hand side can be computed. 

The right-hand side of (\ref{report8}) is different from zero, unlike the case of the equations of motion obtained in \cite{BGKZ2014}. We presently lack a clear understanding of the geometrical meaning of the extra terms, although some insight could be gained by developing the inverse Higgs mechanism to a higher order than that presented in Section \ref{sec_AdS3_geo}.

\section{Geometric interpretation of the general \adst\  action}	
\label{sec_AdS3_geo}	

Let us reconsider the Lagrangian 	\bref{cba_new_26_b}. It is useful to introduce a (partial) parametrization 
of  \adst\ in terms of the coset \adstsod\ and the SO(2) rotations
\be
g=g_{\frac{AdS_3}{SO(2)}} e^{i M_{12}\phi }.
\ee
The MC 1-forms are 	
\begin{eqnarray}
	\TL^0 &=& L^0,\label{gerbertL0}\\
	\TL^{12} &=& L^{12}+\dif\phi,\label{gerbertL12}\\
	\TL^1 &=& L^1 \cos\phi -L^2\sin\phi,\label{gerbertL1}\\
	\TL^2 &=& L^1 \sin\phi +L^2\cos\phi,\label{gerbertL2}\\
	\TL^{01} &=& L^{01} \cos\phi -L^{02}\sin\phi,\label{gerbertL01}\\
	\TL^{02} &=& L^{01}\sin\phi +L^{02}\cos\phi,\label{gerbertL02}
\end{eqnarray} 
where $L^0, L^1, L^2, L^{01}, L^{02}, L^{12}$ are those of \adstsod\ (see \ref{apMC}).

If we disregard the total derivative term in $\TCL^{12}$ in the Lagrangian (\ref{cba_new_26_b})
we get the general \adst\ invariant Lagrangian
\begin{equation}\label{Lcovgen2}
	\CL_{\text{gen}} = - M  \CL^0 - J  \CL^{12} + K_1 \cos\phi + K_2 \sin\phi,
\end{equation}
where
\begin{eqnarray}
	K_1 &=& A_1 \CL^1 + A_2 \CL^2 + B_1 \CL^{01}+ B_2 \CL^{02},\label{gerbertK1}\\
	K_2 &=& -A_1 \CL^2 + A_2 \CL^1 - B_1 \CL^{02}+ B_2 \CL^{01},\label{gerbertK2}
\end{eqnarray}
are independent of $\phi$.
The variable $\phi$ is thus not dynamical, and can be eliminated from its EOM as
$$
\tan\phi=\frac{K_2}{K_1},
$$
which allows to write down the reduced Lagrangian
\begin{equation}\label{Lcovgen3}
	\CL_{\text{gen}}^{\text{red}} =-M \CL^0 - J  \CL^{12}+ \sqrt{K_1^2+K_2^2}. 
\end{equation}
The first two terms in $\CL_{\text{gen}}^{\text{red}}$ correspond to the well known \adstsod\ Lagrangian 
\cite{BGKZ2014}.
The third term of  (\ref{Lcovgen3}) is the new contribution with respect to the \adstsod\ case.
In order to  understand it, we will eliminate the boost variables $v^1$, $v^2$  
by imposing $\CL^1=0$ and $\CL^2=0$, which is a particular case of what is known as the inverse Higgs mechanism \cite{Ivanov-Ogievetskii, Brauner:2014aha}. 
This is equivalent to using the equations of motion for $v^1$, $v^2$ given by $\CL^0$,
\begin{eqnarray}
	\frac{\partial \CL^0}{\partial v^1} &=& - \cosh v^2\ \CL^2,\label{cba_new_30a}\\
	\frac{\partial \CL^0}{\partial v^2} &=& \CL^1.\label{cba_new_30b}
\end{eqnarray}
The mixing of $1\leftrightarrow 2$ is due to $(-v^2)$ being the coordinate associated to the $M_{01}$ generator; see (\ref{cba_new_1}). Solving $\CL^1=0$ and $\CL^2=0$ for $v^1$ and $v^2$  one obtains
\begin{eqnarray}
	\tanh v^1 &=& \frac{\dot x^2}{\dot x^0 \cosh x^1 \cosh x^2},\label{cba_new_31a}\\
	\tanh v^2 &=& -\frac{\dot x^1 \cosh x^2}{\sqrt{(\dot x^0)^2 \cosh^2 x^1 \cosh^2 x^2 - (\dot x^2)^2}}.\label{cba_new_31b}
\end{eqnarray}

Notice that the inverse Higgs mechanism is equivalent to using the EOM for $v^1$, $v^2$ given by $\CL^0$, but not by the full Lagrangian  
$\CL_{\text{gen}}^{\text{red}}$. In this sense, this is the zeroth order computation in the parameters $J$, $A_1$, $A_2$, $B_1$, $B_2$ of the perturbation procedure described in \cite{Gomis:2013NPB} to obtain effective Lagrangians starting with a given seeding Lagrangian.

Using (\ref{cba_new_31a}) and  (\ref{cba_new_31b}) or, equivalently, $\CL^1=0$, $\CL^2=0$, the third term in (\ref{Lcovgen3}) simplifies to
\begin{equation}\label{cba_new_32}
	\sqrt{K_1^2+K_2^2} = \sqrt{B_1^2 + B_2^2} \sqrt{(\CL^{01})^2 + (\CL^{02})^2} = B \sqrt{(\CL^{01})^2 + (\CL^{02})^2},
\end{equation}
where $\CL^{01}$ and $\CL^{02}$ are to be computed with $v^1$, $v^2$ and their time-derivatives expressed in terms of the geometric variables $x^0$, $x^1$, $x^2$ and their first and second order time-derivatives. 

One can then compute
\begin{equation}\label{cba_new_33}
	\sqrt{K_1^2+K_2^2} = B  \sqrt{-g}\  \kappa_1,
\end{equation}
where $\sqrt{-g}$ is the world-line metrics of the curve in $AdS_3$,
\begin{equation}\label{cba_new_34}
	-g = (\dot x^0)^2 \cosh^2 x^1 \cosh^2 x^2 - (\dot x^1)^2 \cosh^2 x^2 - (\dot x^2)^2,
\end{equation}
and $\kappa_1$ is the extrinsic curvature of the world-line,
\begin{equation}\label{cba_new_35}
	\kappa_1^2 = (D_s^2 x)^2
\end{equation}
with $D_s$ the covariant derivative along the world-line \cite{Gomis:2013NPB}
\begin{equation}\label{cba_new_36}
	D_s = \frac{1}{\sqrt{-g}} \frac{\dif}{\dif \tau}
\end{equation}
and $x(s)=(x^0(s),x^1(s),x^2(s))$.

We have thus succeeded in giving an interpretation to the extra terms of our Lagrangian with respect to the one in \cite{BGKZ2014}, but it has been at the cost of using a partial inverse Higgs mechanism. This, in general, changes the dynamics of the remaining variables \cite{Pons:2010JMP},  and the equations of motion of the Lagrangian with the term containing the extrinsic curvature of the world-line are not equivalent to the equations of the original Lagrangian, and in particular to those analyzed in Section \ref{sec_AdS3_lag}.

\section{\adsd\ general Lagrangian}
\label{sec_AdS2_lag}
As stated in the introduction, the dynamical sectors that appear in the general \adst\ Lagrangian can be analyzed by writing it as two copies of an \adsd\ Lagrangian, which we introduce in this section.

We parametrize an element of AdS$_2$, with AdS$_2$ radius $R$, by
\begin{equation}\label{cba6}
	g=e^{iP_0 x^0}e^{iP_1 x^1}e^{i M_{01}y}
\end{equation}
with $P_0$, $P_1$, $M_{01}$ satisfying
\begin{equation}\label{cba7}
	[P_0,P_1]=-i \frac{1}{R^2} M_{01},\quad [M_{01},P_0]=-i P_1,\quad [M_{01},P_1]=-i P_0.
\end{equation}
The mapping with the standard $so(2,1)$ algebra given by the $j_a$,
\begin{equation}\label{cba8}
	[j_0,j_1]=i j_2,\quad [j_2,j_0]=i j_1,\quad [j_2,j_1]=i j_0
\end{equation}
is obtained, for instance, by the identification $ R P_0 \leftrightarrow j_0$, $ R P_1 \leftrightarrow j_1$, $M_{01} \leftrightarrow -j_2$. In this basis one has
\begin{equation}\label{cba6j}
	g=e^{ij_0 X^0}e^{ij_1 X^1}e^{i j_{2} X^2}
\end{equation}
with $ x^0=  R X^0$, $x^1= R X^1$, $y=-X^2$. The MC form in the $j_a$ basis is given by
\begin{eqnarray}
	-i g^{-1} \dif g &=& (\cosh X^2 \cosh X^1\ \dif X^0 + \sinh X^2 \ \dif X^1  ) j_0\nonumber \\
	&+&    (\sinh X^2 \cosh X^1\ \dif X^0 + \cosh X^2\ \dif X^1 ) j_1  \nonumber \\
	&+& (-\sinh X^1\ \dif X^0 + \dif X^2) j_2 \label{ads2_1},        
\end{eqnarray}
and hence the more general AdS$_2$ invariant Lagrangian is
\begin{equation}\label{ads2_2}
	\CL_{\text{gen AdS$_2$}} =  \mu \CL^0 + \alpha \CL^1 + \beta \CL^2,
\end{equation}
with $\mu$, $\alpha$, $\beta$ arbitrary parameters and
\begin{eqnarray}
	\CL^0 &=& \cosh X^2 \cosh X^1 \dot X^0 + \sinh X^2 \dot X^1,\label{ads2_L0}\\
	\CL^1 &=& \sinh X^2 \cosh X^1 \dot X^0 + \cosh X^2 \dot X^1, \label{ads2_L1}\\
	\CL^2 &=& - \sinh X^1 \dot X^0 + \dot X^2. \label{ads2_L2}
\end{eqnarray}

Using (\ref{cba_N3}) and (\ref{cba8}), a generic variation of (\ref{ads2_2}) is given by
\begin{eqnarray}
	\delta\CL_{\text{gen AdS$_2$}} &=&  \mu \frac{\dif}{\dif \tau} [\delta Z]^0  + \alpha \frac{\dif}{\dif \tau} [\delta Z]^1 + \beta \frac{\dif}{\dif \tau} [\delta Z]^2\nonumber\\
	&+& \frac{1}{2} [\delta Z]^0 (-\alpha \CL^2 + \beta \CL^1) + \frac{1}{2} [\delta Z]^1 (-\mu \CL^2 - \beta \CL^0) + \frac{1}{2} [\delta Z]^2 (\mu \CL^1 + \alpha \CL^0).\nonumber\\  
	& &\label{ads2_3}
\end{eqnarray}
The variations 
$\delta[Z]^A$ are given by (see (\ref{ads2_L0})---(\ref{ads2_L2}))
\begin{eqnarray}
	\delta[Z]^0 &=& \cosh X^2 \cosh X^1 \delta X^0 + \sinh X^2 \delta X^1,\label{ads2_dL0}\\
	\delta[Z]^1 &=& \sinh X^2 \cosh X^1 \delta X^0 + \cosh X^2 \delta X^1, \label{ads2_dL1}\\
	\delta[Z]^2 &=& - \sinh X^1 \delta X^0 + \delta X^2, \label{ads2_dL2}
\end{eqnarray} 
and one can see that independent variations of the $X^a$ yield independent $\delta [Z]^A$, so that the EOM are obtained by putting to zero the terms multiplying each $\delta [Z]^A$.
The matrix $S$ in (\ref{cba_N7}) for the case of AdS$_2$ is given by
\begin{equation}\label{ads2_4}
	S=\left(
	\begin{array}{ccc}
		0 & \beta & -\alpha \\ -\beta & 0 & -\mu \\ \alpha & \mu & 0
	\end{array}
	\right),
\end{equation}
which has rank $2$ unless $\alpha=\beta=\mu=0$. The independent EOM are thus any two of $-\alpha \CL^2 + \beta \CL^1=0$, $\mu \CL^2 + \beta \CL^0=0$, $\mu \CL^1 + \alpha \CL^0=0$.  Since the rank can only be 2 or 0 (in which case there is no Lagrangian), no sectors are present in the general \adsd\ Lagrangian and the only gauge transformation is the one corresponding to reparametrizations.
If we disregard the total derivative in $\CL^2$, the variable $X^2$ is non-dynamical, and can be safely removed using its EOM:
\begin{equation}\label{ads2_5}
	\frac{\partial \CL_{\text{gen AdS$_2$}}}{\partial X^2} = \mu \CL^1 + \alpha \CL^0 =0.
\end{equation}
The reduced Lagrangian is then
\begin{equation}\label{ads2_6}
	\delta\CL_{\text{gen AdS$_2$}}^{\text{red}} = \left(  \mu-\frac{\alpha^2}{\mu}       \right) \CL_\text{red}^0 -\beta \sinh X^1 \dot X^0,
\end{equation}
where ${\cal L}^0_{\rm{red}}$ is obtained from ${\cal L}^0$ by replacing $X^2$ with the solution from (\ref{ads2_5}), namely
\begin{equation}\label{ads2_7}
	-\tanh X^2 = \frac{\mu \dot X^1 + \alpha \cosh X^1 \ \dot X^0}{\mu \cosh X^1\ \dot X^0 + \alpha \dot X^1}.
\end{equation}
One finally gets
\begin{equation}\label{ads2_8}
	{\cal L}_{\rm{red}} = \sqrt{(\mu^2-\alpha^2)(\cosh^2 X^1 (\dot X^0)^2 - (\dot X^1)^2)} -\beta \sinh X^1 \dot X^0.
\end{equation}
The first contribution is the standard kinetic term for a particle in AdS$_2$. Notice that, after the elimination of $X^2$, the effect of having $\alpha\neq 0$ is just a redefinition of the mass $\mu$. Also, it follows from  $\frac{\partial \CL^0}{\partial X^2} = \CL^1$ that
the equation of motion for $X^2$ given by ${{\cal L}^0}$ is equivalent to imposing the inverse Higgs mechanism ${\cal L}^1=0$ for $X^2$, a fact that was discussed in \cite{McArthur2010} for a general class of systems in the framework of nonlinear realization theory.

The last term in (\ref{ads2_8}), 
\be\label{ads2_9}
L_{em}= \beta \sinh X^1 \dot X^0,
\ee
corresponds to the interaction of the particle with an electromagnetic background.
Indeed,  this term  can be symmetrized with respect to the velocities by using 
$$
\dot X^0\, \sinh X^1=\frac{1}{2}\left(\dot X^0\, \sinh X^1-X^0 \cosh X^1\dot X^1\right)+\mbox{total derivative},
$$ 
which can be written as $A_a \dot x^a$, with an electromagnetic potential $1$-form
\begin{equation}
	\label{ads2_10}
	A = \frac{1}{2}\sinh X^1\ \dif X^0 -\frac{1}{2}X^0 \cosh X^1\ \dif X^1. 
\end{equation}

The reason why the background does not break any symmetry of \adsd\ is the fact that the field strength is proportional
to the volume form of \adsd,
\begin{equation}
	\label{ads2_11}
	\dif A = - \cosh X^1\ \dif X^0 \dif X^1 = - e^0 \wedge e^1, 
\end{equation}
with  $e^0$, $e^1$ computed from the  Maurer-Cartan form associated to the geometric\footnote{Geometric in the sense that it does not depend on the phase-space variable $X^2$, associated to the boost.} element $g_0=e^{ij_0 X^0}e^{ij_1 X^1}$, which has components   
\begin{equation}\label{cba_new_22}
	e^0=
	\cosh {X}^1\,\dif X^0, \quad e^1=\dif X^1, \quad \w^{01}=\sinh{X}^1\,\dif {X}^0.
\end{equation}

\section{\adsd$\times$\adsd\ Lagrangian}
\label{sec_AdS2xAdS2_lag}

The $so(2,2)$ algebra can be written as the direct sum of two (chiral) $so(2,1)$ algebras with generators   
$j_a^{\pm}$, $a=0,1,2$, related to the (covariant) generators of $so(2,2)$ by
\begin{equation}\label{cba3}
	j_a^{\pm} = \frac{1}{2}\left(-\frac{1}{2}\epsilon_{abc} M^{bc} \pm P_a\right),\quad \epsilon_{012}=-\epsilon^{012}=-1.
\end{equation}
The $j_a^{\pm}$ satisfy
\begin{equation}\label{cba_new_3}
	[j_a^{\pm},j_b^{\pm}] = -i \epsilon_{ab}^{\ \ c}j_{c}^{\pm},
\end{equation} 
with vanishing commutator for any two generators from different chiral sectors.

Using $M_{01}=-j_2^+-j_2^-$, $M_{02}=j_1^++j_1^-$, $M_{12}=j_0^+ + j_0^-$, $P_0=j_0^+-j_0^-$, $P_1=j_1^+-j_1^-$, and $P_2=j_2^+-j_2^-$, the element in (\ref{cba_new_1}) can be rewritten as
\begin{eqnarray}
	g&=&e^{ij^+_0x^0}e^{ij^+_{1}x^{1}}e^{ij^+_{2}x^{2}}e^{ij^+_{1}v^{1}}e^{ij^+_{2}v^{2}}e^{i j^+_0 \phi }
	e^{-ij_0^-x^0}e^{-ij_{1}^-x^{1}}e^{-ij_{2}^-x^{2}}e^{ij_{1}^-v^{1}}e^{ij_{2}^-v^{2}}e^{i j^-_0 \phi }\nonumber\\
	&\equiv& g^+ g^-,\label{cba4}
\end{eqnarray}
with $g^\pm$ elements of \adsd, which we can parametrize as
\bea
g^+&=& e^{ij^+_0X_+^{0}}e^{ij^+_{1}X_+^1 }e^{ij^+_{2}X_+^2},
\nn\\
g^-&=& e^{-ij^-_0X_-^{0}}e^{-ij^-_{1}X_-^1}e^{-ij^-_{2}X_-^2}.
\label{cba5}\eea
The explicit minus signs in the expression for $g^-$ are put for later convenience.

Consider now an element of \adst\ expressed as the product of two  chiral elements of \adsd, as given by (\ref{cba4}) and (\ref{cba5}). One has, using the commutativity between the two chiral factors,
\begin{eqnarray}
	-i g^{-1}\dif g &=& -i((g^-)^{-1}(g^+)^{-1}) ( \dif g^+ g^- + g^+ \dif g^-) = -i (g^+)^{-1} \dif g^+ -i (g^-)^{-1} \dif g^-\nn \\
	&=& L_+^0 j_0^+ + L_+^1 j_1^+ + L_+^{01} j_2^+   +   L_-^0 j_0^- + L_-^1 j_1^- + L_-^{01} j_2^-,
	\label{cba_new_23}
\end{eqnarray}
where the $L_{\pm}^A$, $A=0,1,01$, can be read off from (\ref{ads2_1}), with an extra minus sign for $X_-^a$:
\begin{eqnarray}
	L_+^0 &=& \cosh X^2_+ \cosh X^1_+ \ \dif X_+^0 + \sinh X_+^2 \ \dif X_+^1   ,\label{ads2+L0}\\
	L_+^1 &=& \sinh X^2_+ \cosh X^1_- \ \dif X_+^0 + \cosh X_+^2 \ \dif X_+^1,\label{ads2+L1}\\
	L_+^2 &=& -\sinh X_+^1 \ \dif X_+^0 + \dif X_+^2,\label{ads2+L2}\\
	L_-^0 &=& -\cosh X^2_- \cosh X^1_- \ \dif X_-^0 + \sinh X_-^2 \ \dif X_-^1   ,\label{ads2-L0}\\
	L_-^1 &=& \sinh X^2_- \cosh X^1_+ \ \dif X_-^0 - \cosh X_-^2 \ \dif X_-^1,\label{ads2-L1}\\
	L_-^2 &=& -\sinh X_-^1 \ \dif X_-^0 - \dif X_-^2.\label{ads2-L2}
\end{eqnarray}

If $g$ is expressed using the \adst\ basis, as in (\ref{cba_new_1}), one gets
\begin{eqnarray}
	-i g^{-1}\dif g &=& \TL^0 P_0 + \TL^1 P_1 + \TL^2 P_2 + \TL^{01} M_{01} + \TL^{02} M_{02} + \TL^{12} M_{12}\nonumber\\
	&=& (j_0^+ - j_0^-)\TL^0 + (j_1^+ - j_1^-)\TL^1 + (j_2^+ - j_2^-)\TL^2 \nn \\
	&+& (-j_2^+ - j_2^-)\TL^{01} + (j_1^+ + j_1^-)\TL^{02} + (j_0^+ + j_0^-)\TL^{12}, 
	\label{cba_new_24}
\end{eqnarray}
where the relation between the chiral and covariant generators has been used. 
Comparing with (\ref{cba_new_23}), one sees that the covariant and chiral MC forms are related by
\begin{eqnarray}
	L_{\pm}^0 &=& \pm \TL^0 + \TL^{12},\nn\\
	L_{\pm}^1 &=& \pm \TL^1 + \TL^{02},\nn\\
	L_{\pm}^{01} &=& \pm \TL^2 - \TL^{01}.\label{cba_new_25}
\end{eqnarray}

Consider now two copies of (\ref{ads2_2})
\begin{equation}
	\CL = \mu_+ \CL_+^0 + \mu_- \CL_-^0 + \alpha_+ \CL_+^1 + \alpha_- \CL_-^1 +  \beta_+ \CL_+^2 + \beta_- \CL_-^2.
	\label{carles1}
\end{equation}
This is, in fact, the most general \adst\ invariant Lagrangian 
that can be constructed using the non-linear realization method, \textit{i.e.} (\ref{cba_new_26_b}).
Using (\ref{cba_new_25}) but written for the Lagrangians associated to the forms, one has, as in \cite{BGKZ2014}, 
\begin {eqnarray}
M&=&-\mu_+ + \mu_-,\label{paramM}\\
J&=&-\mu_+ - \mu_-,\label{paramJ}
\end{eqnarray}
together with
\begin{eqnarray}
A_1 &=& \alpha_+-\alpha_-,\label{paramA1}\\
A_2 &=& \beta_+-\beta_- ,\label{paramA2}\\
B_1 &=& -\beta_+ - \beta_- ,\label{paramB1}\\
B_2 &=& \alpha_+ + \alpha_-.\label{paramB2}
\end{eqnarray}

According to the discussion in Section  \ref{sec_AdS3_lag}, one can see from the above relations that the critical cases correspond to the total switching off of one of the two chiral sectors. From this point of view, the extra gauge transformations can be interpreted in terms of the total disappearance of the variables corresponding to one of the   chiral sectors from the Lagrangian.

\section{\adsd$\times$\adsd\ to \adst\ map}
\label{sec_map}

When the coefficients are related by (\ref{paramM}) to (\ref{paramB2}), the Lagrangians (\ref{carles1}) and (\ref{cba_new_26_b}) are the same, but expressed in different coordinates, namely the six chiral ones of \adsd$\times$\adsd\ and the six covariant ones of \adst. The relation between the coordinates is given by (\ref{cba4}) and (\ref{cba5}), namely

\begin{eqnarray}
\lefteqn{
	e^{ij^+_0x^0}e^{ij^+_{1}x^{1}}e^{ij^+_{2}x^{2}}e^{ij^+_{1}v^{1}}e^{ij^+_{2}v^{2}}e^{i j^+_0 \phi }
	e^{-ij_0^-x^0}e^{-ij_{1}^-x^{1}}e^{-ij_{2}^-x^{2}}e^{ij_{1}^-v^{1}}e^{ij_{2}^-v^{2}}e^{i j^-_0 \phi } }\nn \\
& &\quad\quad\quad\quad\quad \quad\quad
=e^{ij^+_0X_+^{0}}e^{ij^+_{1}X_+^1 }e^{ij^+_{2}X_+^2}
e^{-ij^-_0X_-^{0}}e^{-ij^-_{1}X_-^1}e^{-ij^-_{2}X_-^2}.
\label{cba_the_map}
\end{eqnarray}

In order to solve this equation for the chiral coordinates in terms of the covariant ones, we introduce \cite{BGKZ2014} a parameter $t$ such that
\begin{eqnarray}\label{T2_phi}
\lefteqn{
	e^{ij_0^+ x^0} e^{ij_1^+ x^1} e^{ij_2^+ x^2} e^{ij_1^+ v^1 t} e^{ij_2^+ v^2 t}\
	e^{-ij_0^- x^0} e^{-ij_1^- x^1} e^{-ij_2^- x^2} e^{ij_1^- v^1 t} e^{ij_2^- v^2 t}  e^{i (j_0^+ +j_0^-)\phi t}
}\nonumber
\\
& & \quad\quad\quad\quad\quad 
=e^{ij_0^+ X_+^0(t)} e^{ij_1^+ X_+^1(t)} e^{ij_2^+ X_+^2(t)}\
e^{-ij_0^- X_-^0(t)} e^{-ij_1^- X_-^1(t)} e^{-ij_2^- X_-^2(t)}.\nonumber\\
& &
\end{eqnarray}
Equation (\ref{cba_the_map}) is recovered for $t=1$, while for $t=0$ one has
\begin{equation}\label{T3_phi}
X_\pm^0(0)=x^0,\quad X_\pm^1(0)=x^1, \quad X_\pm^2(0)=x^2,
\end{equation}
which will serve as initial conditions for the system of differential equations that we will derive from (\ref{T2_phi}).

Splitting (\ref{T2_phi}) into chiral sectors, one has, for the $+$ sector, 
\begin{equation}\label{T4_phi}
e^{ij_0^+ x^0} e^{ij_1^+ x^1} e^{ij_2^+ x^2} e^{ij_1^+ v^1 t} e^{ij_2^+ v^2 t}e^{i j_0^+\phi t}
= e^{ij_0^+ X_+^0(t)} e^{ij_1^+ X_+^1(t)} e^{ij_2^+ X_+^2(t)}.
\end{equation}
If we write the left-hand side as
$$
g_0^+(t)e^{ij_0^+ t\phi}\equiv g_{AdS_3}^+(t),
$$ 
we get
\begin{eqnarray}
& & -i (g_{AdS_3}^+(t))^{-1}\partial_t g_{AdS_3}^+(t) \nn \\
&=& -i e^{-i j_0^+ t \phi} (g_0^+(t))^{-1} \left( 
\partial_t g_0^+(t) e^{ij_0^+ t \phi} + g_0^+(t) i j_0^+ \phi e^{ij_0^+ t \phi}
\right)\nn\\
&=& e^{-i j_0^+ t \phi} \left(j_2^+ v^2 + j_1^+ v^1 \cosh tv^2 + j_0^+ v^1 \sinh tv^2\right) e^{ij_0^+ t \phi} + j_0^+ \phi.\nn\\
&=& v^2 ( j_2^+ \cos t\phi - j_1^+ \sin t\phi) + v^1 \cosh tv^2 ( j_1^+ \cos t\phi + j_2^+ \sin t\phi) \nn \\
&+& v^1 \sinh tv^2 j_0^+ + \phi j_0^+. 
\end{eqnarray}
The coefficients of $j_0^+$, $j_1^+$ and $j_2^+$ must equal the corresponding coefficients of  
the expression obtained when the left-hand side of  (\ref{T4_phi}) is used for the computation, and one gets
\begin{eqnarray}
\cosh X_+^2 \cosh X_+^1\ \dot X_+^0 + \sinh X_+^2 \ \dot X_+^1 &=& v^1 \sinh tv^2+\phi,\label{T4a_phi}\\
\sinh X_+^2 \cosh X_+^1\ \dot X_+^0 + \cosh X_+^2 \ \dot X_+^1 &=& v^1 \cosh tv^2\cos t\phi- v^2 \sin t\phi,\label{T4b_phi}\\
-\sinh X_+^1\  \dot X_+^0 + \dot X_+^2 &=& v^2\cos t\phi + v^1 \cosh tv^2 \sin t\phi.\label{T4c_phi}
\end{eqnarray}
Solving this for the derivatives yields
\begin{eqnarray}
\dot X_+^0 &=& \frac{\cosh X_+^2}{\cosh X_+^1} ( v^1 \sinh tv^2 + \phi) - \frac{\sinh X_+^2}{\cosh X_+^1} ( v^1 \cosh tv^2 \cos t\phi - v^2 \sin t\phi),\nn\\
\dot X_+^1 &=& \cosh X_+^2 ( v^1 \cosh tv^2\cos t\phi - v^2 \sin t\phi) - \sinh X_+^2 ( v^1 \sinh tv^2 + \phi),\nn\\
\dot X_+^2 &=& \sinh X_+^1 \dot X_+^0 + v^2 \cos t\phi + v^1 \cosh tv^2 \sin t\phi,
\end{eqnarray}
where the first equation must be substituted into the last one. Similarly, for the $-$ sector one gets
\begin{eqnarray}
\dot X_-^0 &=& -\frac{\cosh X_-^2}{\cosh X_-^1} ( v^1 \sinh tv^2 + \phi) - \frac{\sinh X_-^2}{\cosh X_-^1} ( v^1 \cosh tv^2 \cos t\phi - v^2 \sin t\phi),\nn\\
\dot X_-^1 &=& -\cosh X_-^2 ( v^1 \cosh tv^2\cos t\phi - v^2 \sin t\phi) - \sinh X_-^2 ( v^1 \sinh tv^2 + \phi),\nn \\
\dot X_-^2 &=& -\sinh X_+^1 \dot X_-^0 - v^2 \cos t\phi - v^1 \cosh tv^2 \sin t\phi.
\end{eqnarray}

The above set of 6 differential equations, together with the initial conditions $X_{\pm}^0(0)=x^0$,  $X_{\pm}^1(0)=x^1$, $X_{\pm}^2(0)=x^2$, completely define the map from the covariant to the chiral coordinates of \adst, which are obtained for $t=1$. The equations can be solved perturbatively as a series in powers of $v^1$, $v^2$ and $\phi$ (they can be solved exactly for $\phi=0$, which corresponds to the  embedding of \adstsod\ in covariant coordinates into the chiral description of \adst). To first order, one gets
\begin{eqnarray}
X_{\pm}^0 &=& x^0 - v^1 \frac{\sinh x^2}{\cosh x^1} \pm \phi \frac{\sinh x^2}{\cosh x^1},\label{fX0}\\
X_{\pm}^1 &=& x^1 \pm v^1 \cosh x^2 - \phi \sinh x^2, \label{fX1}\\
X_{\pm}^2 &=& x^2 \pm v^2  \mp  v^1 \tanh x^1 \sinh x^2 + \phi \tanh x^1 \cosh x^2.\label{fX2}
\end{eqnarray}

One can check that substituting this into (\ref{carles1}) one obtains the same terms as the Lagrangian (\ref{cba_new_26_b}), using the explicit parametrization given in \ref{apMC}, when computed to first order in $v^1$, $v^2$, $\phi$.

\section{Discussion}
\label{sec_diss}

We have written down and analyzed the most general Lagrangian for a particle in an AdS$_3$ background in the framework of non-linear realizations. Dynamical sectors and gauge transformations have been identified, generalizing the results in\cite{BGKZ2014}. In order to better understand the resulting equations of motion, we have tried to eliminate the non-geometrical variables corresponding to the general Lorentz transformation. Unfortunately, some of them, the ones corresponding to boosts, are dynamical, and their elimination can only be done perturbatively. The final expression, see equation (\ref{report8}), is the equation of the geodesics for a particle in AdS$_3$ but modified by a non-zero right-hand side, whose interpretation is not clear in terms of AdS$_3$ geometry.

We have constructed a general \adsd\ Lagrangian, which, differently to what happens with \adst, has no critical sectors, and has only a gauge transformation corresponding to reparametrizations.
The AdS$_2$ 
Lagrangian contains a non-dinamical variable, whose elimination produces a kinetic term with a redefined mass and a term which can be interpreted as an interaction with an electromagnetic background.

Using the isomorphism   $so(2,1)\times so(2,1)\sim so(2,2)$, we have written an equivalent Lagrangian expressed in terms of two AdS$_2$ Lagrangians.  
From the relation between the parameters of both formulations, it is seen that the critical  AdS$_3$  sectors correspond to the switching off of one of the two chiral sectors,  and hence the extra gauge transformations of the 2 critical sectors of the \adst\ Lagrangian can be understood as  transformations that allow to set to zero the  variables of one of the \adsd\ components.

A differential equation relating the covariant  AdS$_3$  variables and the chiral ones has been obtained. This equation can be solved perturbatively in terms of the variables of the general Lorentz transformation, and we give the results to first order. 

One of the problems that we were not able to address satisfactorily is the geometrical interpretation of the modification of the geodesic equations. The elimination of the   variables  associated to the general Lorentz transformation can be done, in principle, at the level of the Lagrangian or of the equations of motion. The variable associated to spatial rotations is not dynamical and can be eliminated from the Lagrangian, using its equation of motion, but those associated to the boosts are dynamical and their equations of motion can only be used inside the other equations of motion \cite{Pons:2010JMP}. Since the equations cannot be integrated in closed form, only a perturbative procedure, starting with the solution found in  \cite{BGKZ2014}, can be carried out, but the results, even at first order, do not have a clear interpretation. It can be seen that implementing this at the level of the Lagrangian produces 
a term containing the extrinsic curvature of the world-line in the \adst\ background \cite{Gomis:2013NPB}, but this alters the original dynamics of the  
remaining, geometrical variables, and hence corresponds to a different theory.

\section*{Acknowledgments}  The authors would like to thank Kiyoshi Kamimura and Jorge Zanelli for long and fruitful discussions related to this paper's subject. 

The work of CB is partially supported by Project MAFALDA (PID2021-126001OBC31, funded by MCIN/ AEI /10.13039/50110001 1033 and by ``ERDF A way of making Europe''), Project MASHED (TED2021-129927B-I00, funded by MCIN/AEI /10.13039/501100011033 and by the ``European Union Next GenerationEU/PRTR''), and Project ACaPE (20121-SGR-00376, funded by  AGAUR-Generalitat de Catalunya). 

The
research of JG was supported in part by PID2022-136224NB-C21 and by the State Agency for Research of the
Spanish Ministry of Science and Innovation through the Unit of Excellence Mar\'ia de Maeztu 2020-2023 award to
the Institute of Cosmos Sciences (CEX2019- 000918-M).

JG acknowledges the hospitality of the Faculty of Sciences and the Institute of Exact and Natural Sciences of the Universidad Arturo Prat, and the Centro de Estudios Cient\'ificos (CECs) in Valdivia, where this work was completed. The stay was partially funded by SIA 85220027 and CECs.

\section*{ORCID}

\noindent Carles Batlle - \url{https://orcid.org/0000-0002-6088-6187}

\noindent Joaquim Gomis - \url{https://orcid.org/0000-0002-8706-2989}

\clearpage

\appendix

\section{Maurer-Cartan forms for \adstsod\ and \adst}\label{apMC}

If we locally parametrize \adstsod\ as 
\begin{equation}\label{cba_new_1b}
g_{\frac{AdS_3}{SO(2)}}=e^{iP_0 x^0}e^{i P_1 x^1}e^{i P_2 x^2} e^{i M_{02} v^1} e^{-i M_{01} v^2},
\end{equation}
the components of the  Maurer-Cartan  form  have explicit expressions\footnote{We set the anti-de  Sitter radius $R=1$, and restore factors of  $R$, if needed, by appropriate dimensional analysis.} \cite{BGKZ2014}
\begin{eqnarray}
L^0 &=& \cosh x^1\cosh x^2 \cosh v^1 \cosh v^2 \ \dif x^0 \nn\\
&+& \cosh x^2 \sinh v^2 \ \dif x^1 - \sinh v^1 \cosh v^2 \ \dif x^2,\label{cosetL0}\\
L^1 &=&  \cosh x^1\cosh x^2 \cosh v^1 \sinh v^2 \ \dif x^0 \nn\\
&+& \cosh x^2 \cosh v^2 \ \dif x^1 - \sinh v^1 \sinh v^2 \ \dif x^2,\label{cosetL1}\\
L^2 &=& -\cosh x^1 \cosh x^2 \sinh v^1 \ \dif x^0 + \cosh v^1 \ \dif x^2,\label{cosetL2}\\
L^{01} &=& \sinh x^1 \cosh v^1 \ \dif x^0 + \sinh v^1 \sinh x^2 \ \dif x^1 - \dif v^2,\label{cosetL01}\\
L^{02} &=& \left(\cosh x^1 \sinh x^2 \cosh v^2  + \sinh v^1 \sinh v^2 \sinh x^1 \right) \dif x^0\nn\\
&+& \cosh v^1 \sinh x^2 \sinh v^2 \ \dif x^1 + \cosh v^2 \ \dif v^1,\label{cosetL02}\\
L^{12} &=& \left(\sinh x^1 \cosh v^2 \sinh v^1  + \cosh x^1 \sinh x^2 \sinh v^2 \right) \dif x^0\nn\\
&+& \sinh x^2 \cosh v^2 \cosh v^1 \ \dif x^1 + \sinh v^2 \ \dif v^1.\label{cosetL12} 
\end{eqnarray}

This induces a parametrization 
\be
g=g_{\frac{AdS_3}{SO(2)}} e^{i M_{12}\phi }
\ee
of AdS$_3$, with Maurer-Cartan component forms
\begin{eqnarray}
\TL^0 &=& L^0,\label{gerbertL0b}\\
\TL^{12} &=& L^{12}+\dif\phi,\label{gerbertL12b}\\
\TL^1 &=& L^1 \cos\phi -L^2\sin\phi,\label{gerbertL1b}\\
\TL^2 &=& L^1 \sin\phi +L^2\cos\phi,\label{gerbertL2b}\\
\TL^{01} &=& L^{01} \cos\phi -L^{02}\sin\phi,\label{gerbertL01b}\\
\TL^{02} &=& L^{01}\sin\phi +L^{02}\cos\phi.\label{gerbertL02b}
\end{eqnarray}

\section{Symmetry of $\CL_{\text{gen \adsd}}$ under \adsd\ transformations} 
\label{apAdS2}

In the basis given by (\ref{cba7}) the Maurer-Cartan form is
\begin{eqnarray}
-ig^{-1}\dif g &=& \left(\cosh x^1 \cosh y\ \dif x^0 - \sinh y\ \dif x^1 \right)P_0\nn\\
&+&
\left( -\cosh x^1\sinh y\ \dif x^0 + \cosh y\ \dif x^1
\right)P_1\nn \\
&+& \left( 
\sinh x^1\ \dif x^0 + \dif y
\right)M_{01}.
\label{cba10}
\end{eqnarray}
Using the world-line pull-back of this form  one obtains the  Lagrangians
\begin{eqnarray}
{\cal L}^0 &=& \cosh x^1\cosh y\ \dot x^0 - \sinh y\ \dot x^1,\label{cba11_0}\\
{\cal L}^1 &=&  -\cosh x^1\sinh y\ \dot x^0 + \cosh y\ \dot x^1  ,\label{cba11_1}\\
{\cal L}^{01} &=& \sinh x^1\ \dot x^0 + \dot y.\label{cba11_01}
\end{eqnarray}

The AdS$_2$ transformations are determined from
\begin{equation}\label{cba14}
[\delta Z]^0 P_0 + [\delta Z]^1 P_1 + [\delta Z]^{01}M_{01}=\epsilon^0 g^{-1}P_0 g + \epsilon^1 g^{-1}P_1 g + \epsilon^{01} g^{-1}M_{01} g,
\end{equation}

One has
\begin{eqnarray}
g^{-1}P_0 g &=& \cosh x^1\cosh y \ P_0 - \cosh x^1\sinh y\ P_1 + \sinh x^1\ M_{01}
\label{cba16_0}\\
g^{-1}P_1 g &=& (-\cos x^0\sinh y - \sin x^0\sinh x^1 \cosh y)P_0\nn\\
&+& (\cos x^0\cosh y + \sin x^0 \sinh x^1 \sinh y)P_1 - \sin x^0 \cosh x^1\ M_{01}
\label{cba16_1}\\
g^{-1}M_{01} g &=& (\cos x^0 \sinh x^1 \cosh y - \sin x^0 \sinh y)P_0\nn\\
&+& (-\cos x^0 \sinh x^1\sinh y + \sin x^0 \cosh y)P_1 + \cos x^0 \cosh x^1\ M_{01},\nn\\
& & \label{cba16_01}
\end{eqnarray}
and then (\ref{cba14}) can be solved to obtain the AdS$_2$ transformations of the variables $x^0$, $x^1$ and $y$,
\begin{eqnarray}
\delta x^0 &=& \epsilon^0 - \epsilon^1 \sin x^0 \tanh x^1 + \epsilon^{01}\cos x^0 \tanh x^1,\label{cba17a}\\
\delta x^1 &=& \epsilon^1 \cos x^0 + \epsilon^{01}\sin x^0,\label{cba17b}\\
\delta y &=& - \epsilon^1 \sin x^0 \frac{1}{\cosh x^1} + \epsilon^{01}\cos x^0 \frac{1}{\cosh x^1}.\label{cba17c}
\end{eqnarray}
One can check that, under these, ${\cal L}^0$, ${\cal L}^1$ and ${\cal L}^{01}$ are invariant. 
{  Computation of the Noether charges ${\cal Q}_0$, ${\cal Q}_1$ and ${\cal Q}_{01}$ associated to these transformations yields the Casimir value
\begin{equation}
	{\cal Q}_0^2-{\cal Q}_1^2-{\cal Q}_{01}^2 = \mu^2-\alpha^2-\beta^2.
\end{equation}

}

\section{Solving the EOM of ${\cal L}^0$ in  \adsd\ in a fixed gauge}
\label{EOM_L0}

Consider ${\cal L}^0$ with $x^0=t$,
\begin{equation} \label{cba18}
{\cal L}^0_{\text{gf}}= \cosh x^1 \cosh y - \sinh y \ \dot x^1.
\end{equation}
\vskip3mm

One gets two second class primary constraints, $p_y=0$ and $p_1=-\sinh y$. In the reduced phase space $x^1,y$, one has the Dirac bracket
\begin{equation}\label{cba19}
\{x^1,y\}^* = -\frac{1}{\cosh y},
\end{equation}
and the Hamiltonian
\begin{equation}\label{cba20}
H^* = -\cosh x^1\cosh y.
\end{equation}

The EOM are
\begin{eqnarray}
\dot x^1 &=& \{x^1,H^*\}^*=\cosh x^1 \tanh y,\label{cba21a}\\
\dot y &=&\{y,H^*\}^*= -\sinh y.\label{cba21b}
\end{eqnarray}
These can be solved if we define 
\begin{eqnarray}
z&=&\sinh y,\label{cba22a}\\
w &= & \sinh x^1.\label{cba22b}
\end{eqnarray}
One obtains immediately that $z$ is a simple oscillator coordinate, $\ddot z = -z$, from which
\begin{equation}\label{cba23}
z(t) = C_1 \cos t + C_2 \sin t.
\end{equation}
Also, from (\ref{cba21b}), 
$$
w=-\dot y=-\frac{\dif}{\dif t}\arcsinh z = -\frac{1}{\sqrt{1+z^2}}\dot z,
$$
so that
\begin{equation}\label{cba24}
-\sqrt{1+z^2} w = -C_1 \sin t + C_2 \cos t.
\end{equation}
In terms of  $x^1$, $y$, solutions (\ref{cba23}) and (\ref{cba24}) are
\begin{eqnarray}
\sinh y &=& C_1 \cos t + C_2 \sin t,\label{cba25a}\\
-\cosh y \sinh x^1 &=& - C_1 \sin t + C_2 \cos t.\label{cba25b}
\end{eqnarray}

These can be solved for $C_1$ and $C_2$, and one gets the two time-dependent constants of motion
\begin{eqnarray}
J_1 &\equiv & C_1 =\cos t \sinh y + \sin t \cosh y \sinh x^1,\label{cba26a}\\
J_2 &\equiv & C_2 =\sin t \sinh y - \cos t \cosh y \sinh x^1.\label{cba26b}
\end{eqnarray}
They satisfy the algebra
\begin{equation}\label{cba27}
\{J_1,J_2\}^* = -1,
\end{equation} 
and can be combined to yield  the time-independent constant of motion  
\begin{equation}\label{cba28}
J=J_1^2+J_2^2 = \sinh^2 y + \cosh^2 y \sinh^2 x^1.
\end{equation}
One has 
\begin{eqnarray}
\{J_1,J  \}^* &=& -2 J_2,\label{cba29a}\\
\{J_2,J  \}^* &=& 2 J_1.\label{cba29b}
\end{eqnarray}

If instead of $C_1$, $C_2 $ one works with $\alpha$, $\beta$ defined by $C_1=\alpha\cos\beta$, $C_2=\alpha\sin\beta$, one gets the alternative set of constants of motion
\begin{eqnarray}
\tilde J_1 &\equiv&\alpha = \sqrt{z^2+(1+z^2)w^2}=\sqrt{\sinh^2 y + \cosh^2 y \sinh^2 x^1}=\sqrt{J},\label{cba30a}\\
\tilde J_2 &\equiv&\beta = t - \arctan\left(\frac{w(1+z^2)}{z}   \right) = t -\arctan\left(\frac{\sinh x^1}{\tanh y} \right).\label{cba30b}
\end{eqnarray}

In the fixed gauge $x^0=t$, \adsd\ transformations must respect this condition. This can only be done if a re-parametrization is added to the original transformations (computed at $x^0=t$):
\begin{eqnarray}
\delta x^1 &=& \delta_{\text{rig}} x^1 + \epsilon \dot x^1,\label{cba31a}\\
\delta y &=& \delta_{\text{rig}} y + \epsilon \dot y,\label{cba31b}
\end{eqnarray}
where $\delta_{\text{rig}} x^1$, $\delta_{\text{rig}} y$ are given by (\ref{cba17b}), (\ref{cba17c}) with $x^0=t$,
\begin{eqnarray}
\delta x^1 &=& \epsilon^1 \cos t +\epsilon^{01}\sin t + \epsilon \dot x^1,\label{cba31aa}\\
\delta y &=& -\epsilon^1 \sin t \frac{1}{\cosh x^1} + \epsilon^{01} \cos t \frac{1}{\cosh x^1}+ \epsilon \dot y,\label{cba31bb}
\end{eqnarray}
and the reparametrization parameter $\epsilon(t)$  is determined from the condition
\begin{equation}\label{cba32}
\left.\delta (x^0-t)\right|_{x^0=t}=\left.\delta_{\text{rig}} x^0\right|_{x^0=t} + \epsilon(t)=0,
\end{equation}
that is
\begin{eqnarray}\label{cba33}
\epsilon(t)&=&-\left.\delta_{\text{rig}} x^0\right|_{x^0=t}\nn\\
& =& - \epsilon^0 + \epsilon^1 \sin t \tanh x^1 - \epsilon^{01}\cos t \tanh x^1.
\end{eqnarray}
Using this in (\ref{cba31aa}) and (\ref{cba31bb}) one immediately sees that 
${\cal L}^0_{\text{gf}}$ is quasi-invariant:
\begin{eqnarray}
\delta {\cal L}^0_{\text{gf}} &=& \epsilon^0 \frac{\dif}{\dif t}\left(-\cosh x^1 \cosh y +\dot x^1 \sinh y  \right)\nn\\
&+&\epsilon^1 \frac{\dif}{\dif t}\left(- \dot x^1 \sinh y \sin t \tanh x^1 + \cosh y \sin t \sinh x^1 \right)\nn\\
&+&\epsilon^{01} \frac{\dif}{\dif t}\left(\dot x^1 \sinh y \cos t \tanh x^1 - \sinh x^1 \cosh y \cos t  \right).\label{cba34}
\end{eqnarray}

\section{Embedding  of \adstsod\ into  \adsd$\times$\adsd}
\label{embedding}

Here we want to express an element of \adstsod\ as an element of  \adsd$\times$\adsd. This can also be viewed as the zeroth order computation in $\phi$ for the full \adst\ to 
\adsd$\times$\adsd\ map. It turns out that this can be solved in closed form.

In order to obtain the relation between the 6 chiral coordinates and the 5 covariant ones, we introduce, as in the full \adst\ case, a parameter $t\in[0,1]$ such that

\begin{eqnarray}\label{T2}
\lefteqn{
	e^{ij_0^+ x^0} e^{ij_1^+ x^1} e^{ij_2^+ x^2} e^{ij_1^+ v^1 t} e^{ij_2^+ v^2 t}\
	e^{-ij_0^- x^0} e^{-ij_1^- x^1} e^{-ij_2^- x^2} e^{ij_1^- v^1 t} e^{ij_2^- v^2 t}  
}\nonumber \\
&=& \mbox{} e^{ij_0^+ X_+^0(t)} e^{ij_1^+ X_+^1(t)} e^{ij_2^+ X_+^2(t)}\
e^{-ij_0^- X_-^0(t)} e^{-ij_1^- X_-^1(t)} e^{-ij_2^- X_-^2(t)},
\end{eqnarray}
with 
\begin{equation}\label{T3}
X_\pm^0(0)=x^0,\quad X_\pm^1(0)=x^1, \quad X_\pm^2(0)=x^2,
\end{equation}
which are obtained from (\ref{T2}) at $t=0$. 
Equation (\ref{T2}) decouples into $+$ and $-$ sectors, and taking the derivative with respect to the parameter $t$ one gets \cite{BGKZ2014}, for the $+$ sector,
\begin{eqnarray}
\cosh X_+^2 \cosh X_+^1\ \dot X_+^0 + \sinh X_+^2 \ \dot X_+^1 &=& v^1 \sinh(tv^2),\label{T4a}\\
\sinh X_+^2 \cosh X_+^1\ \dot X_+^0 + \cosh X_+^2 \ \dot X_+^1 &=& v^1 \cosh(tv^2),\label{T4b}\\
-\sinh X_+^1\  \dot X_+^0 + \dot X_+^2 &=& v^2.\label{T4c}
\end{eqnarray}

{
The left-hand side of (\ref{T2}) is an element of  \adstsod, so equating it to the right-hand side, which is a general element of \adsd$\times$\adsd, imposes restrictions on the later. In \cite{BGKZ2014} this was solved by setting $X_+^0=X_-^0\equiv X^0$, introducing the coordinate $\alpha$, and interpreting the extra factor as an element of SO(2).}

Equations (\ref{T4a},\ref{T4b},\ref{T4c}) can be solved for the derivatives, and one gets
\begin{eqnarray}
\dot z &=& - v^1 \frac{\sinh y}{\cosh x},\label{T5a}\\
\dot x &=& v^1 \cosh y, \label{T5b}\\
\dot y &=& - v^1 \tanh x \sinh y,\label{T5c}
\end{eqnarray}
where
\begin{equation}\label{T6}
x\equiv X_+^1,\quad y\equiv X_+^2 - t v^2,\quad z\equiv X_+^0.
\end{equation}
From the last two of these one gets
$$
\ddot x = ((v^1)^2-\dot x^2) \tanh x.
$$
It follows that $w=\sinh x$ obeys a simple oscillator equation, $\ddot w = (v^1)^2 w$, from which $w(t)=C_1 e^{v^1 t}+C_2 e^{-v^1 t}$, and hence
\begin{equation}\label{T7}
x(t)=\arcsinh \left(C_1 e^{v^1 t}+C_2 e^{-v^1 t}  \right). 
\end{equation}
Then, from (\ref{T5b}),
$$
\cosh y = \frac{\dot x}{v^1}= \frac{C_1 e^{v^1 t}-C_2 e^{-v^1 t}}{\sqrt{1+\left(   C_1 e^{v^1 t}+C_2 e^{-v^1 t}  \right)^2}    }
$$
and
\begin{equation}
y(t)=\arccosh\left(
\frac{C_1 e^{v^1 t}-C_2 e^{-v^1 t}}{\sqrt{1+\left(   C_1 e^{v^1 t}+C_2 e^{-v^1 t}  \right)^2}    }
\right).
\label{T8}
\end{equation}
The integration constants $C_1$ i $C_2$ can be written in terms of the initial conditions as
\begin{eqnarray}
C_1 &=& \frac{1}{2}\left(\sinh x(0) +\cosh x(0) \cosh y(0)                  \right),\label{T9a}\\
C_2 &=& \frac{1}{2}\left(\sinh x(0) -\cosh x(0) \cosh y(0)                  \right).\label{T9b}
\end{eqnarray}
Using (\ref{T7}) and (\ref{T8}), the last differential equation (\ref{T5b}) becomes
\begin{equation}\label{T10}
\dot z = - v^1 \frac{\sqrt{-4C_1C_2-1}}{1+\left(  C_1 e^{v^1 t}+C_2 e^{-v^1 t} \right)^2}.
\end{equation}
Notice that
$$
-4C_1C_2-1 = \cosh^2 x(0) \sinh^2 y(0) \geq 0.
$$
Integrating (\ref{T10}) one finally gets
\begin{equation}\label{T11}
z(t)=z(0)+\arctan\frac{2C_1^2+2C_1C_2+1}{\sqrt{-4C_1C_2-1}}-\arctan\frac{2C_1^2e^{2v^1t}+2C_1C_2+1}{\sqrt{-4C_1C_2-1}}.
\end{equation}

Setting $t=1$ in (\ref{T7}), (\ref{T8}) and (\ref{T11}), and also in (\ref{T6}), one gets the surjective mapping from the covariant  $(x^0,x^1,x^2,v^1,v^2)$ to the  chiral $(X_+^0,X_+^1,X_+^2)$ coordinates
\begin{eqnarray}
X_+^0 &=& x^0 +\arctan\frac{2C_1^2+2C_1C_2+1}{\sqrt{-4C_1C_2-1}}-\arctan\frac{2C_1^2e^{2v^1}+2C_1C_2+1}{\sqrt{-4C_1C_2-1}},\label{T12a}\\
X_+^1 &=& \arcsinh \left(C_1 e^{v^1} +C_2 e^{-v^1}            \right),\label{T12b}\\
X_+^2 &=& v^2 + \arccosh \frac{C_1 e^{v^1}-C_2 e^{-v^1}}{\sqrt{ 1+\left(  C_1 e^{v^1 }+C_2 e^{-v^1 } \right)^2 }},\label{T12c}
\end{eqnarray}
with
\begin{eqnarray}
C_1 &=& \frac{1}{2}\left(\sinh x^1 + \cosh x^1 \cosh x^2  \right),\label{T13a}\\
C_2 &=& \frac{1}{2}\left(\sinh x^1 - \cosh x^1 \cosh x^2  \right).\label{T13b}
\end{eqnarray}

For the $-$ chiral sector the differential equations of the mapping are given by {(these can also be obtained from the equations in \cite{BGKZ2014} with $\alpha=0$)}
\begin{eqnarray}
-\cosh X_-^2 \cosh X_-^1\ \dot X_-^0 + \sinh X_-^2 \ \dot X_-^1 &=& v^1 \sinh(tv^2),\label{T14a}\\
\sinh X_-^2 \cosh X_-^1\ \dot X_-^0 - \cosh X_+^2 \ \dot X_-^1 &=& v^1 \cosh(tv^2),\label{T14b}\\
-\sinh X_-^1\  \dot X_-^0 -\dot X_-^2 &=& v^2.\label{T14c}
\end{eqnarray}
and they can be mapped to those of the $+$ sector by $X_-^1\mapsto -X_+^1$ and $v^2\mapsto -v^2$, together with $x^1\mapsto -x^1$  in order to maintain the initial condition $X_-^1(0)=x^1$. Under $x^1\to -x^1$  one can see that $C_1 \to -C_2$ and $C_2\to - C_1$, and hence we can straightforwardly obtain the mapping for the $-$ chiral sector from the one for the $+$ sector:
\begin{eqnarray}
X_-^0 &=& x^0 +\arctan\frac{2C_2^2+2C_1C_2+1}{\sqrt{-4C_1C_2-1}}-\arctan\frac{2C_2^2e^{2v^1}+2C_1C_2+1}{\sqrt{-4C_1C_2-1}},\label{T15a}\\
X_-^1 &=& \arcsinh \left(C_2 e^{v^1} +C_1 e^{-v^1}            \right),\label{T15b}\\
X_-^2 &=& -v^2 + \arccosh \frac{-C_2 e^{v^1}+C_1 e^{-v^1}}{\sqrt{ 1+\left(  C_2 e^{v^1 }+C_1 e^{-v^1 } \right)^2 }}.\label{T15c}
\end{eqnarray}

The part of the complete   \adsd$\times$\adsd Lagrangian  proportional to the \adsd\ world-line lengths is (eq. (17) in \cite{BGKZ2014}, but with $X_+^0$ and $X_-^0$)
\begin{eqnarray}
{\cal L}^{\text{ch}} &=& 
\mu^+ (\cosh X_+^2\cosh X_+^1 \dot X_+^0 + \sinh X_+^2 \dot X_+^1 )\nn\\
&+&
\mu^- (-\cosh X_-^2\cosh X_-^1 \dot X_-^0 + \sinh X_-^2 \dot X_-^1 ).
\label{T16}
\end{eqnarray}

Using $\mu^\pm=-\frac{1}{2}(J\pm M) $ and  (\ref{T12a})---(\ref{T12c}), (\ref{T15a})---(\ref{T15c}) one can obtain the Lagrangian in covariant coordinates. For instance, setting $J=M$ one has
$$
{\cal L}^+ = -M (\cosh X_+^2\cosh X_+^1 \dot X_+^0 + \sinh X_+^2 \dot X_+^1 ).
$$
Using   (\ref{T12a})---(\ref{T12c}) one can see that 
\begin{eqnarray}
\lefteqn{-\frac{1}{M}{\cal L}^+ =}\nn\\ & &\mkern-18mu  \left( \cosh v^2 \cosh v^1 \cosh x^2 \cosh x^1 + \sinh v^2 \sinh x^2 \cosh x^1 + \cosh v^2 \sinh v^1 \sinh x^1 \right)\dot x^0\nn \\
&+& \left( \sinh v^2 \cosh x^2 + \cosh v^2 \cosh v^1 \sinh x^2  \right)\dot x^1 \nn \\
&-& \cosh v^2 \sinh v^1 \dot x^2  + \sinh v^2 \dot v^1,\nn\\\label{T17}
\end{eqnarray}
which coincides with the \adstsod\  Lagrangian in the critical case $M=J$. 

A key intermediate result in the proof of this is
\begin{eqnarray}\label{T18}
\cosh X_+^2 \cosh X_-^1 
& =& \cosh v^2 \sinh x^1 \sinh v^1 + \cosh v^2 \cosh x^1 \cosh x^2 \cosh v^1 \nn\\ &+& \sinh v^2 \cosh x^1 \sinh x^2,
\end{eqnarray}
as well as combining the two $\arctan$ in $X_+^0$ to obtain
\begin{equation}\label{T19}
X_+^0 = x^0 - \arctan \frac{\sinh x^2 \sinh v^1 }{\sinh x^1 \cosh x^2 \sinh v^1 + \cosh x^1 \cosh v^1}.
\end{equation}

Similarly, one can put $J=-M$ in (\ref{T16}) and, using (\ref{T15a})---(\ref{T15c}), obtain the \adstsod\  Lagrangian in the critical case $M=-J$, that is
\begin{eqnarray}
\lefteqn{-\frac{1}{M}{\cal L}^- =}\nn\\ & &\mkern-18mu  \left( \cosh v^2 \cosh v^1 \cosh x^2 \cosh x^1 - \sinh v^2 \sinh x^2 \cosh x^1 - \cosh v^2 \sinh v^1 \sinh x^1 \right)\dot x^0\nn \\
&+& \left( \sinh v^2 \cosh x^2 - \cosh v^2 \cosh v^1 \sinh x^2  \right)\dot x^1 
- \cosh v^2 \sinh v^1 \dot x^2  - \sinh v^2 \dot v^1.\nn\\ \label{T17minus}
\end{eqnarray}

By combining the results of both critical cases, one sees that the transformation of the sum of the two 
$\mu^{+} {\cal L}_+^0$ and $\mu^{-} {\cal L}_-^0$ 
chiral Lagrangians yields the complete $M+J$ covariant Lagrangian in the coset \adstsod. 

\bibliographystyle{JHEP}
\bibliography{AdS3}

\providecommand{\href}[2]{#2}\begingroup\raggedright\begin{thebibliography}{10}

\bibitem{Maldacena:1997re}
J.M.~Maldacena, \emph{{The Large N limit of superconformal field theories and
  supergravity}}, \href{https://doi.org/10.4310/ATMP.1998.v2.n2.a1}{\emph{Adv.
  Theor. Math. Phys.} {\bfseries 2} (1998) 231}
  [\href{https://arxiv.org/abs/hep-th/9711200}{{\ttfamily hep-th/9711200}}].

\bibitem{Gubser:1998bc}
S.S.~Gubser, I.R.~Klebanov and A.M.~Polyakov, \emph{{Gauge theory correlators
  from noncritical string theory}},
  \href{https://doi.org/10.1016/S0370-2693(98)00377-3}{\emph{Phys. Lett. B}
  {\bfseries 428} (1998) 105}
  [\href{https://arxiv.org/abs/hep-th/9802109}{{\ttfamily hep-th/9802109}}].

\bibitem{Witten:1998qj}
E.~Witten, \emph{{Anti-de Sitter space and holography}},
  \href{https://doi.org/10.4310/ATMP.1998.v2.n2.a2}{\emph{Adv. Theor. Math.
  Phys.} {\bfseries 2} (1998) 253}
  [\href{https://arxiv.org/abs/hep-th/9802150}{{\ttfamily hep-th/9802150}}].

\bibitem{Aharony:1999ti}
O.~Aharony, S.S.~Gubser, J.M.~Maldacena, H.~Ooguri and Y.~Oz, \emph{{Large N
  field theories, string theory and gravity}},
  \href{https://doi.org/10.1016/S0370-1573(99)00083-6}{\emph{Phys. Rept.}
  {\bfseries 323} (2000) 183}
  [\href{https://arxiv.org/abs/hep-th/9905111}{{\ttfamily hep-th/9905111}}].

\bibitem{Hubeny:2014bla}
V.E.~Hubeny, \emph{{The AdS/CFT Correspondence}},
  \href{https://doi.org/10.1088/0264-9381/32/12/124010}{\emph{Class. Quant.
  Grav.} {\bfseries 32} (2015) 124010}
  [\href{https://arxiv.org/abs/1501.00007}{{\ttfamily 1501.00007}}].

\bibitem{Gibbons:2011sg}
G.W.~Gibbons, \emph{{Anti-de-Sitter spacetime and its uses}},  in \emph{{2nd
  Samos Meeting on Cosmology, Geometry and Relativity: Mathematical and Quantum
  Aspects of Relativity and Cosmology}}, pp.~102--142, 10, 2011
  [\href{https://arxiv.org/abs/1110.1206}{{\ttfamily 1110.1206}}].

\bibitem{Kraus:2006wn}
P.~Kraus, \emph{{Lectures on black holes and the AdS(3) / CFT(2)
  correspondence}}, {\emph{Lect. Notes Phys.} {\bfseries 755} (2008) 193}
  [\href{https://arxiv.org/abs/hep-th/0609074}{{\ttfamily hep-th/0609074}}].

\bibitem{Martinec:2023zha}
E.J.~Martinec, \emph{{AdS$_{3}$ orbifolds, BTZ black holes, and holography}},
  \href{https://doi.org/10.1007/JHEP10(2023)016}{\emph{JHEP} {\bfseries 10}
  (2023) 016} [\href{https://arxiv.org/abs/2307.02559}{{\ttfamily
  2307.02559}}].

\bibitem{PhysRevD.48.1506}
M.~Ba\~nados, M.~Henneaux, C.~Teitelboim and J.~Zanelli, \emph{Geometry of the
  2+1 black hole}, \href{https://doi.org/10.1103/PhysRevD.48.1506}{\emph{Phys.
  Rev. D} {\bfseries 48} (1993) 1506}.

\bibitem{PhysRevD.79.105011}
O.~Mi\ifmmode \check{s}\else \v{s}\fi{}kovi\ifmmode~\acute{c}\else \'{c}\fi{}
  and J.~Zanelli, \emph{Negative spectrum of the $2+1$ black hole},
  \href{https://doi.org/10.1103/PhysRevD.79.105011}{\emph{Phys. Rev. D}
  {\bfseries 79} (2009) 105011}.

\bibitem{Alvarez:2007fw}
P.D.~Alvarez, J.~Gomis, K.~Kamimura and M.S.~Plyushchay, \emph{(2+1)d exotic
  newton–hooke symmetry, duality and projective phase},
  \href{https://doi.org/https://doi.org/10.1016/j.aop.2007.03.002}{\emph{Annals
  of Physics} {\bfseries 322} (2007) 1556}.

\bibitem{BGKZ2014}
C.~Batlle, J.~Gomis, K.~Kamimura and J.~Zanelli, \emph{Dynamical sectors for a
  spinning particle in {${\mathrm{AdS}}_{3}$}},
  \href{https://doi.org/10.1103/PhysRevD.90.065017}{\emph{Phys. Rev. D}
  {\bfseries 90} (2014) 065017}.

\bibitem{Skagerstam:1989ti}
B.-S.~Skagerstam and A.~Stern, \emph{Topological quantum mechanics in 2+1
  dimensions},
  \href{https://doi.org/10.1142/S0217751X90000714}{\emph{International Journal
  of Modern Physics A} {\bfseries 05} (1990) 1575}
  [\href{https://arxiv.org/abs/https://doi.org/10.1142/S0217751X90000714}{{\ttfamily
  https://doi.org/10.1142/S0217751X90000714}}].

\bibitem{deSousaGerbert:1990yp}
P.~{de Sousa Gerbert}, \emph{On spin and (quantum) gravity in 2 + 1
  dimensions},
  \href{https://doi.org/https://doi.org/10.1016/0550-3213(90)90288-O}{\emph{Nuclear
  Physics B} {\bfseries 346} (1990) 440}.

\bibitem{deAzcarraga:1982dw}
J.A.~de~Azcarraga and J.~Lukierski, \emph{{Supersymmetric Particles with
  Internal Symmetries and Central Charges}},
  \href{https://doi.org/10.1016/0370-2693(82)90417-8}{\emph{Phys. Lett. B}
  {\bfseries 113} (1982) 170}.

\bibitem{Siegel:1983hh}
W.~Siegel, \emph{Hidden local supersymmetry in the supersymmetric particle
  action},
  \href{https://doi.org/https://doi.org/10.1016/0370-2693(83)90924-3}{\emph{Physics
  Letters B} {\bfseries 128} (1983) 397}.

\bibitem{MZ}
O.~Mi\ifmmode \check{s}\else \v{s}\fi{}kovi\ifmmode~\acute{c}\else \'{c}\fi{}
  and J.~Zanelli, \emph{Negative spectrum of the $2+1$ black hole},
  \href{https://doi.org/10.1103/PhysRevD.79.105011}{\emph{Phys. Rev. D}
  {\bfseries 79} (2009) 105011}.

\bibitem{Gomis:2006xw}
J.~Gomis, K.~Kamimura and P.C.~West, \emph{{The construction of brane and
  superbrane actions using non-linear realisations}},
  \href{https://doi.org/10.1088/0264-9381/23/24/010}{\emph{Class. Quant. Grav.}
  {\bfseries 23} (2006) 7369}
  [\href{https://arxiv.org/abs/hep-th/0607057}{{\ttfamily hep-th/0607057}}].

\bibitem{Gomis:2013NPB}
J.~Gomis, K.~Kamimura and J.M.~Pons, \emph{{Non-linear realizations, Goldstone
  bosons of broken Lorentz rotations and effective actions for p-branes}},
  \href{https://doi.org/10.1016/j.nuclphysb.2013.02.018}{\emph{Nuclear Physics
  B} {\bfseries 871} (2013) 420–451}.

\bibitem{Bergshoeff:2022eog}
E.~Bergshoeff, J.~Figueroa-O'Farrill and J.~Gomis, \emph{{A non-lorentzian
  primer}},
  \href{https://doi.org/10.21468/SciPostPhysLectNotes.69}{\emph{SciPost Phys.
  Lect. Notes} {\bfseries 69} (2023) 1}
  [\href{https://arxiv.org/abs/2206.12177}{{\ttfamily 2206.12177}}].

\bibitem{Anabalon:2006ii}
A.~Anabalón, J.~Gomis, K.~Kamimura and J.~Zanelli, \emph{N=4 superconformal
  mechanics as a non linear realization},
  \href{https://doi.org/10.1088/1126-6708/2006/10/068}{\emph{Journal of High
  Energy Physics} {\bfseries 2006} (2006) 068–068}.

\bibitem{PhysRev.166.1568}
S.~Weinberg, \emph{Nonlinear realizations of chiral symmetry},
  \href{https://doi.org/10.1103/PhysRev.166.1568}{\emph{Phys. Rev.} {\bfseries
  166} (1968) 1568}.

\bibitem{PhysRev.177.2239}
S.~Coleman, J.~Wess and B.~Zumino, \emph{Structure of phenomenological
  lagrangians. i}, \href{https://doi.org/10.1103/PhysRev.177.2239}{\emph{Phys.
  Rev.} {\bfseries 177} (1969) 2239}.

\bibitem{BANDOS199277}
I.A.~Bandos and A.A.~Zhelukhin, \emph{Green-schwarz superstrings in spinor
  moving frame formalism},
  \href{https://doi.org/https://doi.org/10.1016/0370-2693(92)91957-B}{\emph{Physics
  Letters B} {\bfseries 288} (1992) 77}.

\bibitem{Bandos:1995zw}
I.A.~Bandos, D.P.~Sorokin, M.~Tonin, P.~Pasti and D.V.~Volkov,
  \emph{{Superstrings and supermembranes in the doubly supersymmetric
  geometrical approach}},
  \href{https://doi.org/10.1016/0550-3213(95)00267-V}{\emph{Nucl. Phys. B}
  {\bfseries 446} (1995) 79}
  [\href{https://arxiv.org/abs/hep-th/9501113}{{\ttfamily hep-th/9501113}}].

\bibitem{West:2000hr}
P.C.~West, \emph{{Automorphisms, nonlinear realizations and branes}},
  \href{https://doi.org/10.1088/1126-6708/2000/02/024}{\emph{JHEP} {\bfseries
  02} (2000) 024} [\href{https://arxiv.org/abs/hep-th/0001216}{{\ttfamily
  hep-th/0001216}}].

\bibitem{Ivanov-Ogievetskii}
E.A.~Ivanov and V.I.~Ogievetskii, \emph{Inverse higgs effect in nonlinear
  realizations}, \href{https://doi.org/10.1007/BF01028947}{\emph{Theoretical
  and Mathematical Physics} {\bfseries 25} (1975) 1050}.

\bibitem{Brauner:2014aha}
T.~Brauner and H.~Watanabe, \emph{{Spontaneous breaking of spacetime symmetries
  and the inverse Higgs effect}},
  \href{https://doi.org/10.1103/PhysRevD.89.085004}{\emph{Phys. Rev. D}
  {\bfseries 89} (2014) 085004}
  [\href{https://arxiv.org/abs/1401.5596}{{\ttfamily 1401.5596}}].

\bibitem{Pons:2010JMP}
J.M.~Pons, \emph{Substituting fields within the action: Consistency issues and
  some applications}, {\emph{Journal of Mathematical Physics} {\bfseries 51}
  (2009) 122903}.

\bibitem{McArthur2010}
I.N.~McArthur, \emph{{Nonlinear realizations of symmetries and unphysical
  Goldstone bosons}},
  \href{https://doi.org/10.1007/JHEP11(2010)140}{\emph{JHEP} {\bfseries 11}
  (2010) 140} [\href{https://arxiv.org/abs/1009.3696}{{\ttfamily 1009.3696}}].

\end{thebibliography}\endgroup

\end{document}